\documentclass[12pt]{article}
\usepackage{amsfonts}
\usepackage{amssymb}
\usepackage[usenames]{color}
\usepackage{epsfig}

\begin{document}
\begin{center}

{\Large\color{Blue} P. Grinevich, A. Mironov, S. Novikov \footnote{P. Grinevich,
Landau Institute, Moscow, e-mail pgg@landau.ac.ru, A. Mironov, Sobolev Math Institute,
Novosibirsk, and Novosibirsk State University,
e-mail mironov@math.nsc.ru, S. Novikov, University of Maryland, College Park
and  Landau/Steklov  Institutes of RAS, Moscow, e-mail novikov@umd.edu,

 Homepage www.mi.ras.ru/$\tilde{}snovikov$
(click publications), the items  177, 178, 179}}

{\Large\color{Blue} Nonrelativistic 2D Purely Magnetic Supersymmetric Pauli
Operator and Solitons}

\end{center}

{\Large Abstract.} {\it The Complete  Manifold  of  Ground State  Eigenfunctions
 for the
Purely Magnetic 2D Pauli Operator is considered as a by-product of the new reduction
 found by the present authors few years ago
for the Algebrogeometric Inverse Spectral Data (i.e. Riemann Surfaces and Divisors).
 This reduction is associated
with the (2+1) Soliton Hierarhy containing a  2D analog of the famous ''Burgers System''.
 This article contains also exposition of the previous works  made since
  1980 including the first topological ideas in the space of quasimomenta.
  We present here also  new results  dedicated to the self-adjoint
 boundary
 problems for Pauli Operator. The  2D zero level ''nonspectral'' Bloch-Floquet functions give discrete
  points of additional spectrum similar to the ''boundary states'' of
  finite-gap 1D potentials in the gaps.}

\vspace{2cm}

{\LARGE Contents}\\
1. Introduction\\
2. History: Ground states, Magnetic Bloch functions. Topology of the
typical dispersion relations.\\
3. Supersymmetry and Laplace transformation.\\
4. Algebro-Geometric scalar Schr\"odinger operators. Reduction problem.\\
5. Spin 1/2. Purely magnetic reduction.\\
6. Solutions of genus zero. Types of real solutions.\\
7. Solutions of genus 1, $\delta$-terms. \\
8. Extension of results to the infinite trigonometric series for $g=0$.\\
9. The Boundary Problems.\\
10. Bloch functions outside of the spectrum as solutions of the
boundary problems.

\pagebreak


{\bf 1. The Introduction.} In general, the motion of charged particle  in the external
electromagnetic field  is described by the full
4D Dirac equation (for spin 1/2).
 Pauli derived a beautiful nonrelativistic approximation $L^P$ of this operator (see \cite{LL}, paragraph 33).
We are going to consider a special 2D case where the electric field is equal to zero, in which the remarkable internal symmetry of this operator is revealed .
The theory of ground states for this operator starts in 1979--80. In the present work we describe
 its unification with the Algebro--Geometric Spectral Theory for one energy level
 of periodic scalar Schr\"odinger operators and corresponding theory of soliton nonlinear systems.
  Especially  analog of the
 2D Burgers hierarchy plays a key role here. The Aharonov--Bohm type terms in magnetic field
 like the ''quantized'' delta function play important role in our construction. This work extends results
 published in \cite{GMN1, GMN2, GMN3, GMN4, GMN5}. We pay special attention to the self-adjoint boundary problems
 allowing to reveal a meaning
 of the  Bloch functions with nonunitary multipliers and another singularities in the theory of quantum states for domains with boundaries.
Extensive varieties of such solutions are found by us.
Similar ideas in the theory of finite-gap Schr\"odinger operators, where the zeros of Bloch functions leads to their interpretations as
``border states'', were known for $D=1$ since 1974 , but
 for $D>1$ there extension never was discussed. {\bf What self-adjoint boundary conditions are realized in quantum mechanics of real systems,
 except Dirichlet?} We do not know an answer to this question.

\vspace{1cm}

{\bf 2. History: Ground states, Magnetic Bloch functions. Topology of the
typical dispersion relations.}

\vspace{0.3cm}

 In  1979--1980 three groups of authors
 studied the ground level  using ''the Factorization Property'' of
the 2D purely Magnetic Pauli Operator written in the Lorenz gauge $A_{1x}+A_{2y}=0$
with $A_1=i\Phi_y,A_2=-i\Phi_x$: Avron--Seiler [AS], \cite{AS},
Aharonov--Casher [AC], \cite{AC}, Dubrovin--Novikov [DN] \cite{DN1, DN2}.

 The Pauli Operator is a second order
2-component Schr\"odinger-type operator. We use a special system of units such that

$$
 L^P=L^+\bigoplus L^-,\quad -L^{\pm}=(\partial_x+i\Phi_y)^2+(\partial_y-i\Phi_x)^2\pm
  \Delta\Phi.
$$
The operator acts on the space of vector-functions $\Psi=(\Psi^+,\Psi^-)$.\\ We assume that
 the charge is equal to 1.
 Following formal observation is extremely useful:

 \newtheorem{lem}{Lemma}
 \begin{lem}
 Let $Q=\partial_z-A_z,\ Q^+=-(\partial_{\bar z}+A_{\bar{z}})$.
 The Scalar Operators $L^{\pm}$ are
 Strongly Factorized   $L^+=QQ^+$, $L^-=Q^+Q$,
   $\partial_z=\partial=\partial_x-i\partial_y, A_z=A_1-iA_2$,
    magnetic field $B=\Delta \Phi$.
\end{lem}
The proof can be easily obtained by the direct substitution.

 The most interesting classes of magnetic fields are [AC] and [DN] bellow.
 Following results were obtained (see proof below):

 1. AC: Rapidly
 decreasing fields,\\ $|[B]|=|\int\int_{{\mathbb R}^2}Bdxdy|<\infty$.
 The ground states form a finite-dimensional space of dimension
 $m\in {\mathbb Z},\\ m\leq [B]/2\pi <m+1$.

 2. DN: Arbitrary periodic fields with integer-valued flux
through the elementary
 cell $\frac{1}{2\pi}\int\int_{cell} Bdxdy=m\in {\mathbb Z}$.\\ The ground states generate an
 infinite dimensional subspace in the Hilbert Space  $L_2({\mathbb R}^2)$
 isomorphic to the Landau level. The so-called ''Magnetic-Bloch states'', eigen for so called magnetic translation (below),  form
  a manifold isomorphic to the total space of the holomorphic vector bundle over
   2-torus of magnetic quasimomentum (below). The projection on base
   exactly coincides with the so-called   ''magnetic
  quasimomentum''. The first Chern class of this bundle is equal to the basic cohomology class in the group
   $H^2(T^2,{\mathbb Z})$, the fibre is isomorphic to ${\mathbb C}^m$. We give corresponding formulas below.

The experience of Differential Topology after Whitney, Pontryagin and
 Thom led S.P. Novikov about 1980 to the idea to study  ''generic'' operators satisfying
 transversality requirements for spectral manifolds, in order to define their
 topological invariants. The fact is that  classes of Schr\"odinger operators
 in quantum theory of particles and crystals are periodic
 and rapidly decreasing, not random. Physical spectra are  characterized not by abstract sets
  and measures,
  but smooth and complex manifolds for which the ideology of differential geometry
 is natural.  It allowed to reveal topological invariants of generic operators
  in the space of quasimomenta. In particular
  Chern numbers of the  dispersion relations appear: Every dispersion relation  is
   complex line bundle
  over the torus $T^2$ in the case of generic position in two-dimensional
  case. They were invented and studied in 1980--81 by S.P. Novikov and
   A.S. Lyskova (see \cite{N1, L1, L2, N2}).
 Schr\"odinger operator in periodic fields (magnetic field must has integer flux through elementary cell)
  is a family of complex Hermitian
elliptic operators over the torus, and its  spectrum is discrete. In generic case spectrum is
 never multiple in the two-dimensional case:  the condition of multiplicity has codimension 3
  among complex Hermitian matrices (or elliptic operators with discrete spectrum on compact manifolds)).
This fact was observed  in the discussion of von Neumann and Wigner in
1930s, but topological consequences were firstly discussed
by Novikov in 1980. Thereby every mode defines a vector line bundle over the 2-torus.
 It was shown, that the first Chern class can be arbitrary here.
This result easily follows from the perturbation theory of homogeneous magnetic field
 with the flux $m>1$ through the elementary cell. Magnetic--Bloch functions
of one Landau level form a vector bundle with the class $c_1\in H^2(T^2,{\mathbb Z})$ which
 is a generator in this group. After  perturbation this
bundle splits into $m$ line sub-bundles with Chern classes $c_1^j$, where $\sum_jc_1^j=c_1$.
 Lyskova showed, that in the rest the terms are
arbitrary and any topological possibility can be realized. In the typical one-parameter families
of two-dimensional Schr\"odinger operators some
isolated points might appear, where two dispersion relations ''collide'', and one hands off to another
 the unity of the Chern class...
   This set of ideas was rediscovered
  by  physicists of the Thouless group  few years later to explain
   the famous ''Integer Quantum
  Hall Fenomenon'' experimentally discovered at that time (\cite{T1, T2}).

\newtheorem{thm}{Theorem}
\begin{thm}
 In the cases AC and DN  all ground states are the
Instantons belonging to one spin-sector only. It means precisely following.\\
 a. They satisfy to the 1st order equations
$Q^+\psi=0$ in the sector $L^+$ for the case $[B]>0$ and $Q\psi=0$ in the sector $L^-$
for the case $[B]<0$.
It is a simple prototype of the self-duality equation.\\
b. They belong to the Hilbert Space $L_2({\mathbb R}^2)$.
\end{thm}
 {\it Proof.} a. The case AC.
Let $\Psi$ is a zero mode for the operator $L^{+}\Psi=0$ and
$\Psi\in L_2$. The standard ''Instanton argument'' is
$$0=<QQ^+\Psi,\Psi>=<Q^+\Psi,Q^+\Psi>$$
for the ground states.
 So $Q^+\Psi=0$. Every square-integrable
solution
of this equation $\Psi\in L_2$ defines a zero mode of the operator $L^+$ and vice versa.
So our result follows if we will be able to find solutions from the Hilbert Space
$L_2({\mathbb R}^2)$.
 Let $0<[B]<\infty,\ m\leq \frac{1}{2\pi} [B]<m+1$ in the case AC (above). We look for $\Psi$ as a product
 of two factors: one is an arbitrary
holomorphic polynomial $P_l,l=0,...,m$, other one depends on magnetic field:
$$\Psi_l=P_l(z)\exp\{- R(x,y)\}$$
with special solution to the equation  $\Delta R=B$ if $[B]>0$:
$$
 R(x,y)=-\frac{1}{4\pi i}\int\int_{{\mathbb R}^2}\ln |z-w|dw\wedge d\bar{w}.
$$
 The growth of $R$ obviously depends on the value
 of magnetic flux $[B]$:
$$
 \int\int_{{\mathbb R}^2}  |\Psi_l|^2dx\wedge dy<\infty
$$
for $l=0,...,m-1$. So we found all ground states in the case AC. There are no square
integrable solutions in the second sector $L^-$.\\
b. The case DN. Solutions here also can be found in one spin-sector only.  As above,
 we assume that the magnetic flux through elementary cell is a
positive integer $m\in Z_+$.
 We consider rectangular lattice with periods $2\omega\in {\mathbb R},2i\omega'\in i{\mathbb R}$.
Let us define $R$ by the formula
$$
 R=-\frac{1}{4\pi i}\int\int_{K}\ln |\sigma (z-w)|B(w,\bar{w})dw\wedge
 d\bar{w}
$$
where $\Delta R=B$, $K$ is elementary cell. We look for the solutions written in the form
$$
 Q^+\Psi=0,\Psi=\exp\{-R(x,y)\}[e^{az}\prod_{j=1,...,m}\sigma(z-a_j)]\lambda
$$
where $\lambda\neq 0$ is any number.

\begin{lem} These solutions are the Magnetic Bloch functions, i.e. the eigenvectors of
magnetic translations $T_j^*,j=1,2$:
$$
 T_1^*\Psi(x,y)=\Psi(x+2\omega,y) e^{-if_1(x,y)}=\varkappa_1\Psi(x,y),
$$
$$
 T_2^*\Psi(x,y)=\Psi(x,y+2\omega')e^{-if_2(x,y)}=\varkappa_2\Psi(x,y).
$$
Vector potential is chosen in the Lorentz gauge form $A=(i\Phi_y,-i\Phi_x)$
for $B=\Delta \Phi$ and $\Phi=-R$ as above. By definition,
$$
 A(x+2\omega,y)=A(x,y)+i(f_{1x}, f_{1y}),\quad A(x,y+2\omega')=A(x,y)+i(f_{2x},f_{2y}).
$$
\end{lem}

{\it Proof.} We use the standard transformation properties
$$
 \sigma(w+2\omega)=-e^{-2\eta
 (w+\omega)}\sigma(w),\quad \sigma(w+2i\omega')=-e^{-2i\eta'(w+i\omega')}\sigma(w)
$$
where $\zeta(w)=\sigma_w/\sigma,\eta=\zeta(\omega),i\eta'=\zeta(i\omega')$ for the
 rectangular lattice. Our theorem immediately follows from this properties by the
 direct calculation. After calculations, we are coming to the following formula for
 such values of parameters  that both multipliers are unitary $|\varkappa_1|=|\varkappa_2|=1$:
$$
 {\rm Re} a={\rm Re}\{\eta_1/\omega[2\sum_ja_j-1/\pi\int\int_KzB(x,y)dxdy]\},
$$
$$
 {\rm Im} a={\rm Im}\{\eta'/\omega'[2\sum_ja_j-1/\pi\int\int_KzB(x,y)dxdy]\}.
$$
By definition, the components of quasimomenta are $p_1,p_2$:
$$
 e^{2ip_1\omega}=\varkappa_1,\quad e^{2ip_2\omega'}=\varkappa_2.
$$
They have a form following from the formulas for the magnetic Bloch functions above:
$$
 p_1+m\pi/2\omega={\rm Im} (a-\eta/\omega\sum_ja_j),
$$
$$
 p_2+m\pi/2\omega'={\rm Re} (a-\eta'/\omega'\sum_ja_j).
$$

The quasimomentum map is well-defined on the space of parameters describing the states
with unitary multipliers. It looks like
$$
 p_1+ip_2=p(a_1,...,a_m)=-(2\pi i/|K|)\sum_ja_j+const
$$
where $|K|=4\omega\omega'$ is an area of the elementary cell $K$.
Its image is a torus $T^2$, the fibre is a complex space ${\mathbb C}^m$.

The elementary arguments presented in the work \cite{DN1, DN2} show that this family describes
all eigenstates of the operator $L^P$ nearby of the ground level $\epsilon=0$.
It follows from the properties of the elliptic operator $L^P$ with fixed unitary multipliers.
Its Kernel is exactly $m$-dimensional because the index of  operator $Q^+$ with
$Q^+:\Psi^+\rightarrow \Psi^-$ is equal to $m$, and the adjoint operator $Q:\Psi^-\rightarrow
\Psi^+$ has no Kernel. No singularities appear in this family for all points in the torus of
 quasimomenta (i.e. for all unitary multipliers). So we conclude that there is a finite nonzero
  gap $\Delta$ separating zero level from all other levels for the operator $L^P$.

\vspace{1cm}

{\bf 3. Supersymmetry and Laplace Transformations.}

\vspace{1cm}

   The operator $S=Q^+:\Psi^+\rightarrow\Psi^-$ and $S:\Psi^-\rightarrow 0$ is
  called  {\bf Super-Symmetry}  for $L^P$. Here $S^2=0,$ $SL^P=L^PS,$ $S^*L^P=L^PS^*$.
  The ''adjoint'' supersymmetry operator is $S^*=Q:\Psi^-\rightarrow \Psi^+$,\\
  $S^*(\Psi^+)=0$. We have
   $SS^*+S^*S=L^P$.\\
   It  implies that all higher
  levels are 2-degenerate  (the ground level is $\infty$-degenerate). The Super-Symmetry
  operator was discussed in physical literature of 1980s (see, e.g. \cite{Fiz}).

  It is in fact we meet hire with a partial case of the so-called
   Laplace Transformations known since XVIII Century for the scalar 2D Schr\"odinger operators.
   Every 2nd order hyperbolic operator
   $L=\partial_x\partial_y+A\partial_x+B\partial_y+V$
   can be presented in the form $L=(\partial_x+B)(\partial_y+A)+W=QQ'+W$. We define
$$
 L\rightarrow
 \tilde{L}=WQ'W^{-1}Q+W,\quad \Psi\rightarrow\tilde{\Psi}=Q'\Psi.
$$
We conclude that $L\Psi=0$ implies $\tilde{L}\tilde{\Psi}=0$ and
$$
 \tilde{L}=(\partial_y+A-W_y/W)
 (\partial_x+B)+W=(\partial_x+B)(\partial_y+A-W_y/W)+\tilde{W}
$$
where
$\tilde{W}=W-A_x+B_y+(\ln W)_{xy}.$

We are going to deal with Elliptic Case important for the applications in Quantum Mechanics, and replace  $\partial_x$ by $\partial=\partial_z$ and
 $\partial_y$ by $\bar{\partial}$.
So we have $L=QQ^++W\rightarrow \tilde{L}=WQ^+W^{-1}Q+W$ and
$\Psi\rightarrow  Q^+\Psi=\tilde{\Psi}$
with $Q=\partial+A,Q^+=-\bar{\partial}+\bar{A}$. {\bf For the special case of
factorizable operators
$W=const$ this transformation acts on the whole spectrum of operator $L$. Physicists identify
 it with
''supersymmetry'' for the Pauli Operator:}
$L\Psi=\epsilon\Psi$
implies $\tilde{L}\tilde{\Psi}=\epsilon\tilde{\Psi}$ for $\tilde{\Psi}=Q^+\Psi$.

The gauge invariants of operator $L=QQ^++W$ are Magnetic field $2B=A_{\bar{z}}-\bar{A}_z$ and
Potential $W$.
It is convenient to write Laplace transformations in terms of invariants only:
$$\tilde{W}=W+\tilde{B},\ \tilde{B}=B+1/2\Delta\log W.$$

It was observed that the whole infinite chain of Laplace Transformations $W_n=e^{f_n}=\tilde{W}_{n-1}, B_n=\tilde{B}_{n-1}$ is equivalent to the
''2D Toda Lattice System''. This observation (of course in hyperbolic case) is going back to the XIX Century.
It was done in the school of Darboux (see \cite{Dar}).
This school also started to consider the Cyclic Chains of Laplace Transformations and
pointed out that they lead to several completely integrable systems including $\sin$-gordon. These works
contains only formal calculations dedicated to the hyperbolic case (see historic discussion in \cite{NV}).
We can exclude the magnetic field from the relation
$W_{n+1}-W_n=B_{n+1}$. It leads to the equation
$$
 1/2\Delta f_n=e^{f_{n+1}}-2e^{f_n}+e^{f_{n-1}}.
$$
The substitution $f_n=g_n-g_{n-1}$ leads this equation to the two-dimensional Toda chain.
 In the work of S.P. Novikov and A.P. Veselov (1997, \cite{NV}) cyclic chains were considered in the elliptic case where
   global restrictions play fundamental role.
  For smooth elliptic double-periodic operators  it was proved in \cite{NV} that
  corresponding Schr\"odinger
   operators $L$ are always Algebro--Geometric, i.e. the collection of nonsingular
    complex Bloch--Floquet
   solutions to the equation $L\Psi=0$ form an algebraic curve (i.e. Riemann surface of finite genus).

The proof of this theorem is based on the fact that solutions of the cyclic chain satisfy an
elliptic partial equation. Thereby periodic nonsingular solutions form finite-dimension varieties.
In complete analogy with arguments of the papers by Novikov and Dubrovin 1974--75 \cite{DMN}, from hire
it follows immediately that this solutions are algebro-geometric, since the cyclic chain is completely integrable
system in the sense of modern theory (Lax representation etc.). Higher flows are linear dependence on the finite-dimensional
variety. This leads to the finite-gap (algebro-geometric) solutions. It seems that the similar argument
 was first noticed by Hitchin \cite{Hi} for harmonic tori in $S^3$ and used by Pinkal and Sterling for studying
 constant mean curvature surfaces of toric topology in ${\mathbb R}^3$ (see survey of Bobenko \cite{Bo})

However, this theory contains only Topologically Trivial operators where all coefficients
 are double periodic, and magnetic flux through the elementary cell is equal to zero.
 In order to include topologically nontrivial operators, another type of Laplace chains was
 considered in \cite{NV}. Let $L_0\rightarrow L_1\rightarrow...\rightarrow L_n$
  be a Laplace transformation chain
 such that the end operators $L_0, L_n$ are both strongly factorizable up to constant:
$$
 L_n=C_n+Q_nQ^+_n,\quad L_0=Q_0Q^+_0, \quad C_n=n[B_0], \quad  [B_0]=2\pi m>0, m\in {\mathbb Z},
$$
where $[B]$ is a flux of the magnetic field $B$ through elementary cell.
 The ground level is highly degenerate here for the smooth double periodic magnetic field $B_0$.
For $n=2$ nontrivial cases were found such that the field $B_n$ is also smooth. Therefore one higher level of energy $\epsilon=n[B_0]$
(originating from the ground level for the operator $Q_nQ^+_n$, thrown to the operator $L_0$ with the help of Laplace chain) is also
infinitely degenerate. To find smooth double periodic solutions of the equation
 $$1/4\Delta u=-e^{u}+[B_0], \quad e^u=W_0=B_0.$$
There are many such solutions. We do not know nontrivial solutions for lengths of chain greater than two ($n>2$).

According to our conjecture, no more than 2 levels may be infinitely degenerate except the case of Landau operator
in the homogeneous magnetic field, and one of them probably is ground.

\vspace{1cm}

{\bf 4. Algebro--Geometric scalar Schr\"odinger operators. Reduction problem.}

\vspace{1cm}

{\bf Question:} Is  Theory of Ground States for the Purely Magnetic
Pauli Operator related somehow to the Algebro--Geometric Theory of
the scalar periodic 2D Schr\"odinger Operators based on the Selected
Energy Level and 2D Soliton Hierarchies?

Let us remind here the history of this subject.

The Algebro--Geometric Theory of the 2D
Second Order  Scalar Schr\"odinger Operators and Corresponding Soliton
Hierarchies based on the selected energy level was started in 1976
by Manakov \cite{M} and Dubrovin, Krichever, Novikov \cite{DKN}.\\
The magnetic field is always Topologically Trivial in this theory (i.e. coefficients of
vector-potentials  are double periodic).
The idea to use the system
$$
 dL/dt=(LH-HL)+fL
$$
instead of ordinary Lax pairs was invented in \cite{M}.
Nontrivial example with both operators $L,H$  of  order 2 was
 demonstrated showing that this new approach is nonempty. Long time this specific system
 was not studied. A ''2-point scalar
 Baker--Akhiezer function''
 was invented in \cite{DKN} for the solution of inverse problems for $L$ based on one energy
  level. It allows to
 construct the
  whole hierarchy of higher systems associated with the original  one.
   In particular, the whole hierarchy
 is completely different from the KP-type situation: it depends on 2
  infinite sets of times $t'_j,t''_j$ associated with 2 ''infinite'' points.
   Our Riemann surface here
  is a family of all nonsingular complex Bloch--Floquet functions of one selected energy level
  $L\psi=0$ ({\bf ''the Complex Fermi Curve''}). In the Algebro--Geometric Case
  it has finite genus. Two ''infinite'' points with local parameters are selected on this algebraic Fermi curve.
The divisor of poles of degree equals to the genus should be specified. This data set defines Bloch function and the operator
which is generally complex and also defines the whole hierarchy of nonlinear systems associated with it
(its deformations with the help of Manakov triples --- above).

 Some reductions in these equations (i.e. dynamically invariant subsystems or sub-hierarchies)
  were actively
studied in 1980s. Several authors found them  either for Nonlinear
Systems or for Inverse Spectral (Scattering)  Data (or for both). Solution of
this problem for the Inverse Spectral Data is  more difficult:
It implies  in particular the description of all reduced hierarchy. However, practically
the existence of reductions is much easier to observe on the level of equation, than for the Inverse Spectral Data.

Our Main Goal  here is Quantum Mechanics and Spectral
Theory. We use Nonlinear Systems as a tool for this goal.

In the first paper 1976 the authors found only purely real reduction of the spectral data.

1. The Data leading to the Self-Adjoint
Periodic   Operators were found by Cherednik in 1980 \cite{Ch}.

2. The  Data leading to the operators $L=-\Delta+U(x,y)$
(with  zero magnetic field)
were found in 1984 (see \cite{NV1, NV2, NV3, N3}). The necessity of Novikov--Veselov conditions has not been proven yet, however
it is certainly correct.

Extension of these results to the
Rapidly Decreasing
Potentials was studied in the  works of S. Manakov, P. Grinevich,
R. Novikov and S. Novikov in  the late 1980s (see \cite{GM, GrN, GN}). Krichever in \cite{K}
    proved that every 2D smooth
    periodic potential can be approximated by the AG ones.

\vspace{1cm}

{\bf 5. Spin 1/2. Purely magnetic reduction.}

\vspace{1cm}

How to find algebro-geometric data for the reduction which leads to the factorized operators?
The Problem was solved recently by authors (see \cite{GMN1, GMN2, GMN3, GMN4, GMN5}).

\vspace{0.2cm}

{\bf Solution of the Reduction Problem}.

\vspace{0.2cm}

Consider a very first Manakov's  System $L_t=[H,L]+fL$ where both  $L,H$ are second order
 operators (1976):
$L=\partial_x\partial_y+G\partial_y+S,\ H=\Delta+F\partial_y+A$,
$$
 G_t=G_{xx}-G_{yy}+(F^2/4)_x-(G^2)_x-A_x+2S_y,
$$
$$
 S_t=S_{yy}-S_{xx}-2(GS)_x+(FS)_y,
$$
with differential constraints
$$
 F_x=2G_y,\quad A_y=2S_x,\quad f=2G_x-F_y.
$$
Konopelchenko already pointed out in 1988 (see \cite{Ko}) that the reduction $S=0$ is
    dynamically invariant for this system. It looks like 2D analog of the famous Burgers system.\\
     We believe in the following {\bf Informal Principle}:
      Every dynamically invariant reduction
     of  completely integrable Soliton system can be effectively
      described in terms of the scattering (spectral) data and Riemann surfaces.\\
   How  to describe this specific reduction  for the
    Inverse Spectral Data?   Is it possible?  Making replacement $x,y\rightarrow
      z,\bar{z}$ we are coming to elliptic operators most interesting for us.

{\bf Our recent result \cite{GMN1,GMN2,GMN3,GMN4,GMN5} describes corresponding Inverse
Problem Data for the operator $L$
with $S=0$}:
Take Riemann Surface (the Complex  Fermi Curve) splitted into
nonsingular pieces $\Gamma=\Gamma'\bigcup\Gamma''$  with  genuses
$g',g''$. They cross each other $P_j=Q_j$, $P_j\in \Gamma'', Q_j\in \Gamma'$, $j=0,...,l$.
(see Fig 1 for the self-adjoint elliptic case where $g'=g''$).
 Take infinities $\infty_1\in
\Gamma',\infty_2\in\Gamma''$ with local parameters
$k'^{-1},k''^{-1}$. Construct function $\psi=(\psi',\psi'')$ with
asymptotic $\psi'\sim c(x,y)e^{k'\bar{z}}(1+O(k'^{-1}))$,
$\psi''=e^{k''z}(1+O(k''^{-1}))$ and    divisors of poles
$D',D''$ of degree $g'+l, g''$  not crossing
infinities and intersection points. In the hyperbolic case take the variables $x,y$
instead of $z,\bar{z}$. Time dynamics can be added in the standard way:
$$
 \psi'=e^{k'\bar{z}+\sum_{s>1}(k')^st'_s}(c+O(k'^{-1})),\ \ \psi''=e^{k''z+\sum_{s>1}(k'')^st''_s}(1+O(k''^{-1})).
$$
The crossing property has a form
$$
 \psi'(Q_j)=\psi''(P_j),\quad j=0,1,...,l.
$$

 \begin{thm} Such Data generate a 2-point Baker--Akhiezer  function $\psi=(\psi',\psi'')$
 on the splitted surface $\Gamma=\Gamma'\bigcup\Gamma''$
 and scalar  operator
 $L'=\Delta+G\partial_{\bar{z}}$
  with $S=0$ such that $L'\psi'=L'\psi''=0$.
  \end{thm}

 {\bf The Data for Self-Adjoint Operators.}
 For the selection of self-adjoint case we need to have $g'=g''$ and to add the degenerate
  Cherednik-type
 restriction. One can say that it is simply a limit of Cherednik condition for the
 degenerate Riemann Surface
 $\Gamma=\Gamma'\bigcup\Gamma''$. We take infinite points $\infty'\in\Gamma'$ with local
 parameter $k'^{-1}$ and $\infty''\in\Gamma''$ with local parameter $k''^{-1}$. An
  anti-holomorphic
 map should be given $\tau:\Gamma''\rightarrow \Gamma'$ such that $\tau(\infty'')=\infty'$,
 $\tau^*(k')=\bar{k''}$, and $\tau$ permutes the intersection points $\tau(P_j)=Q_{k_j}$.
 The inverse map $\Gamma'\rightarrow\Gamma''$ we also call $\tau$, so we have
  $\tau:\Gamma\rightarrow\Gamma,\tau^2=1$.
The divisor of poles is $D'\in\Gamma'$ and $D''\in\Gamma''$ with degrees $|D'|=g'+l,|D''|=g''$
where $g'=g''$ and poles do not cross the intersection points and infinities. The total divisor
 of poles at the unified Riemann Surface $\Gamma$ is $D=D'\bigcup D''$.
  We require as usually that there exists a meromorphic one-form $\Omega$ with simple zeroes
   in the points
  of divisors $D,\tau(D)$ and simple poles in the infinite points $\infty_1,\infty_2$.
  The divisor $(K)$
  consists of  the zeroes of any holomorphic one-form, $\sim$ is a ''linear equivalence'' of
  divisors:
 $$D+\tau(D)\sim (K)+\infty'+\infty''.$$
 Now let our surface be degenerate. The one-form $\Omega$
 also became degenerate. Its degeneration consists of two meromorphic one-forms $\omega',\omega''$
 in the Riemann
 surfaces $\Gamma',\Gamma''$ with simple poles in the crossing points $P_j\rightarrow \tau(Q_j)$
 such that ${\rm res}_{P_j}\omega''+{\rm res}_{Q_j}\omega'=0$. Their divisors should satisfy to the
 conditions:
$$
 (\omega')=D'+\tau(D'')-\infty'+\sum_j Q_j,\quad
 (\omega'')=D''+\tau(D')-\infty''+\sum_j P_j.
$$

  After the non-unitary gauge transformation
$$
 L'\rightarrow L=\frac{1}{\sqrt{c}}L'\sqrt{c},\quad \psi\rightarrow \frac{\psi}{\sqrt{c}}
$$  in $\Gamma',\Gamma''$,
 we obtain a self-adjoint operator
$$
 L=QQ^+=(\partial-A)(-\bar{\partial}-\bar{A}),\quad A=\frac{1}{2}\bar{\partial}\log c.
$$

 Taking $L^+=L $ and $L^-=Q^+Q$, we construct a Purely Magnetic Pauli Operator\\
$L^P=QQ^+\bigoplus Q^+Q$. The Magnetic Field is real  $B=1/2\Delta\log c$,
periodic or quasiperiodic and {\bf topologically trivial}.
It is nonsingular if $c\neq 0$, so the operator is self-adjoint in this case.

 {\bf How to find  ground states in the Hilbert space?}
It is very simple in the periodic nonsingular case $c\neq 0$.
  Take $\psi_0=c^{1/2}$ in
the first spin-sector $L^+$ because $Q^+\psi_0=0$. \\Take $\phi_0=c^{-1/2} $
in the second sector $L^-$
because $Q\phi_0=0$.  {\bf In the case of periodic smooth real $c\neq 0$
 we have exactly two periodic ground state functions  located in both sectors. They
  present the bottom of the CONTINUOUS SPECTRUM near the ground level $\epsilon=0$.}
 See below full description of all complex nonsingular Bloch functions of the ground level.
  Their relationship to the magnetic Bloch functions
 found in \cite{DN1} in 1980 for the topologically nontrivial
 magnetic field also will be discussed below.

\vspace{1cm}

{\bf 6. Solutions of genus zero. Types of real solutions.}

\vspace{0.3cm}

 \begin{center}
 {\color{Black} Fig 1}\\
 \vspace{5mm}
\mbox{\epsfysize=12cm\epsffile{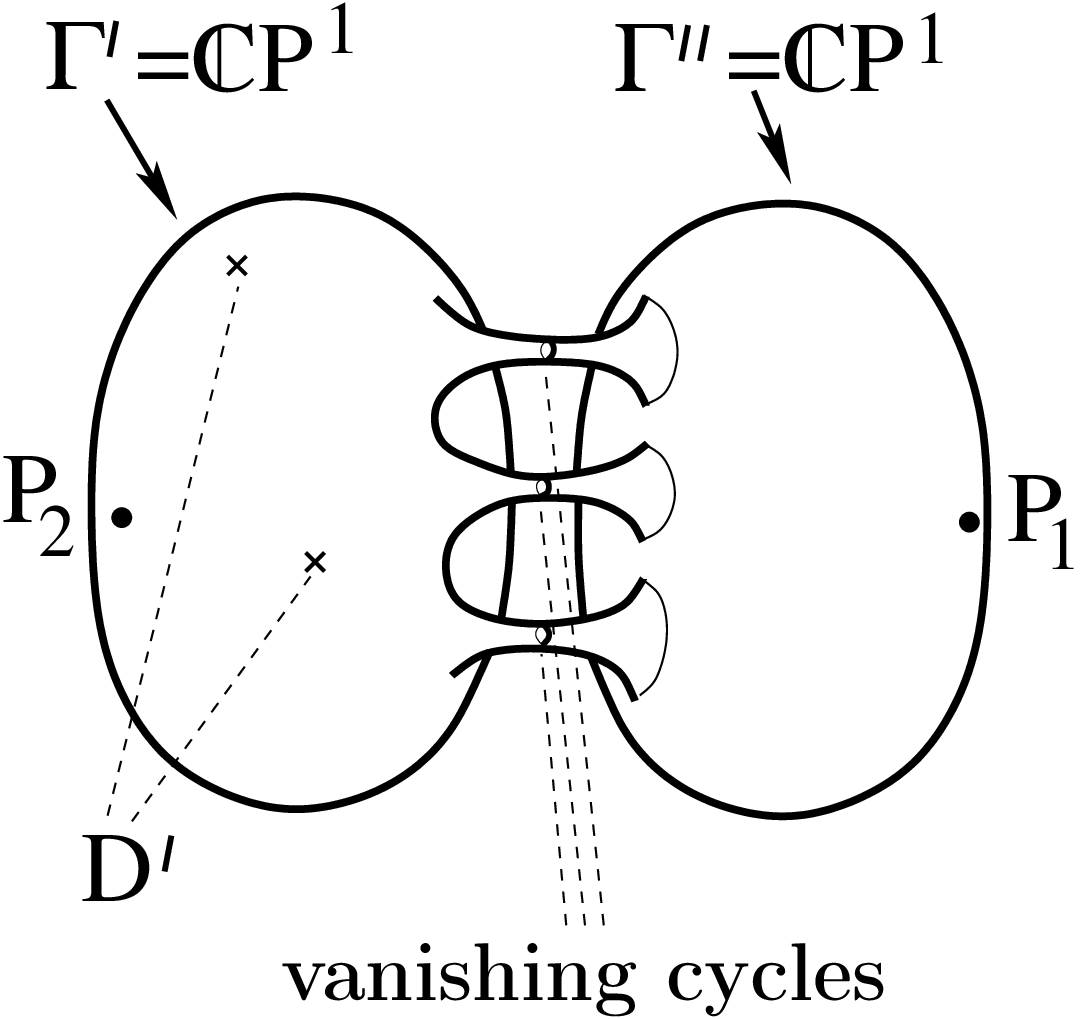}}
\end{center}

 We take $l+1$ intersection points
presented as $k'=k_s$ and $k''=p_s$ in $\Gamma',\Gamma''$,
and divisor $D'=(a_1,...,a_l)$ of degree $l$ in $\Gamma'$. These algebro-geometric data defines Baker--Akhiezer function. We have
$$
 \Psi'=e^{k'\bar{z}}\frac{w_0k'^l+...+w_l}{(k'-a_1)...(k'-a_l)},\
 \Psi'|_{k'=k_s}=e^{p_sz}.
$$
As we can see, $c=w_0$.
So  $c=\sum_{s=0}^{l}\kappa_se^{W_s(z,\bar{z})}$,
 where $W_s=p_sz-k_s\bar{z}$ (see \cite{GMN2}).
In this paper the following important theorem is proved:

\begin{thm}
 For  every trigonometric/exponential polynomial $c$ of the form above with
 any set of linear forms $W_s$
 and complex coefficients $\kappa_s$  there exists AG Data generating such function $c$.
\end{thm}

Let $ W_s=\alpha_sx+\beta_sy,(\alpha_s,\beta_s)\in C^2_W$.
\\Transformation
 $c\rightarrow c'=ce^{\gamma+\alpha x+\beta y}$
 leads to the gauge equivalent operator (with the same magnetic field).

\vspace{1cm}

{\bf  There exist 3 types of Real Solutions:}\\
1. Purely Exponential Positive Case (``The Lump-type fields'')\\
   $\kappa_s>0,(\alpha_s,\beta_s)\in {\mathbb R}^2$.\\
2. Periodic Trigonometric Real
   Case. It will be considered below jointly with the case $g=1$.\\
3. Mixed  exponential/trigonometric case. It can be realized only if
  its ''dominating part'' belongs to the case 1. So we will not discuss it. \\

 The case 1. Let
  ''the Tropical Sum'' of the forms in the set $\{W\}$ is
  nonnegative $I'_{\{W\}}(\phi)=
 \max_s(\alpha_s\cos\phi+\beta_s\sin\phi)\geq 0$.\\
  Then  $c^{-1/2}$ is bounded in ${\mathbb R}^2$.\\
  For the angles $I'_{\{W\}}(\phi)>0$ we have a rapid decay \\
  $c^{-1/2}\rightarrow 0,R\rightarrow\infty$,
  Let $I(\phi)=\max\{I'(\phi),0\}$.

\pagebreak

  \begin{center}
  {\color{Black} Fig 2a}\\
 \vspace{1cm}
  \mbox{\epsfysize=8cm\epsffile{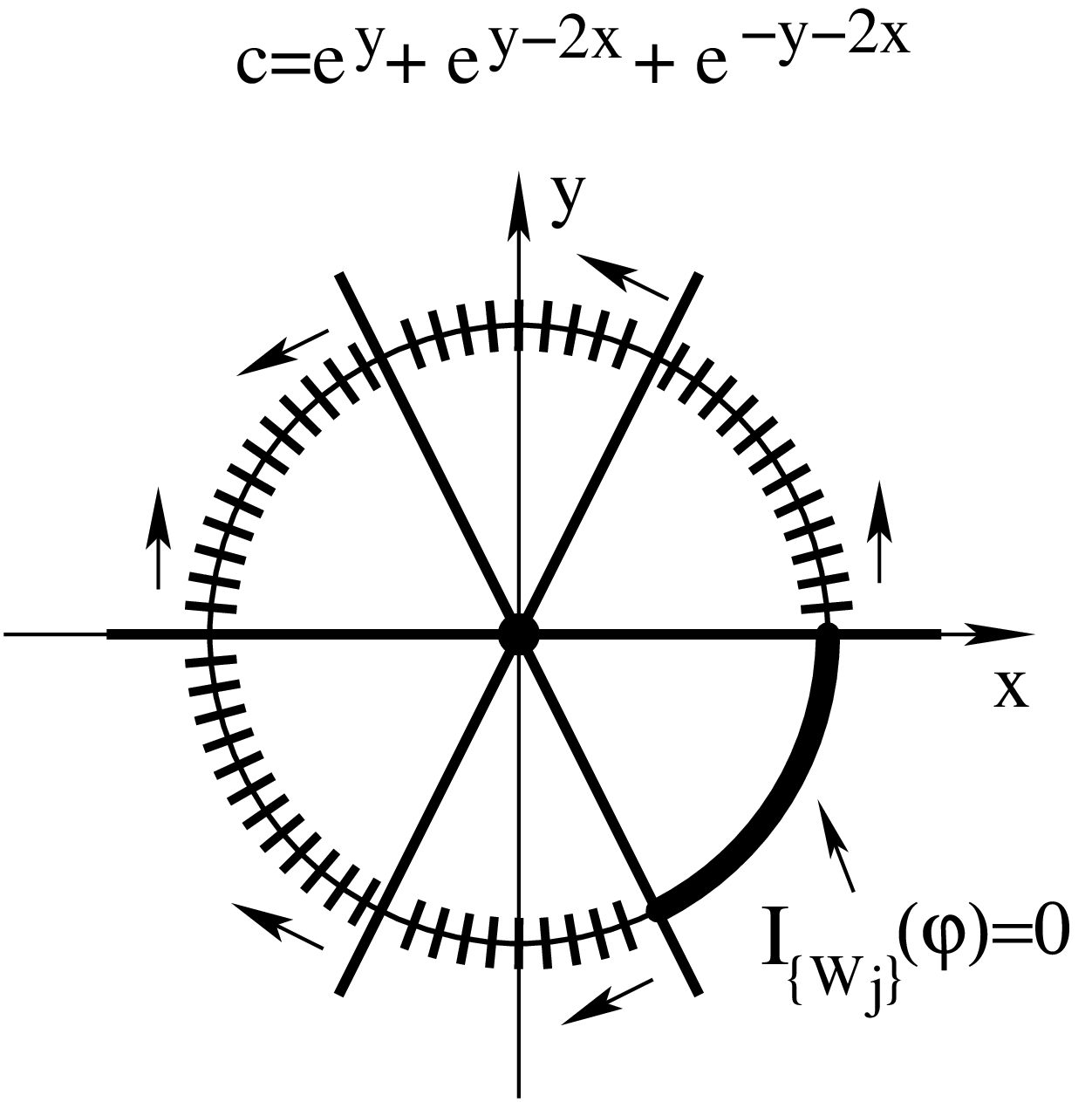}}
  \end{center}

In every class $c'\in {ce^W},
W'\in R^2_W$, the set of representatives $c'$ with nonnegative
$I=I'_{\{W'\}}(\phi)\geq 0$ forms a convex polytop $\bar{T_c}$. Its
inner part $T_c\subset\bar{T_c}$ consists of all $c'$ such that
$I_{\{W'\}}>0$. Open part $T_c$ is always nonempty for $l>2$.
$\bar{T_c}$ is nonempty for $l>1$  (see Fig 2b for $l=3$).

\pagebreak

\begin{center}
{\color{Black} Fig 2b}\\

 \mbox{\epsfysize=8cm\epsffile{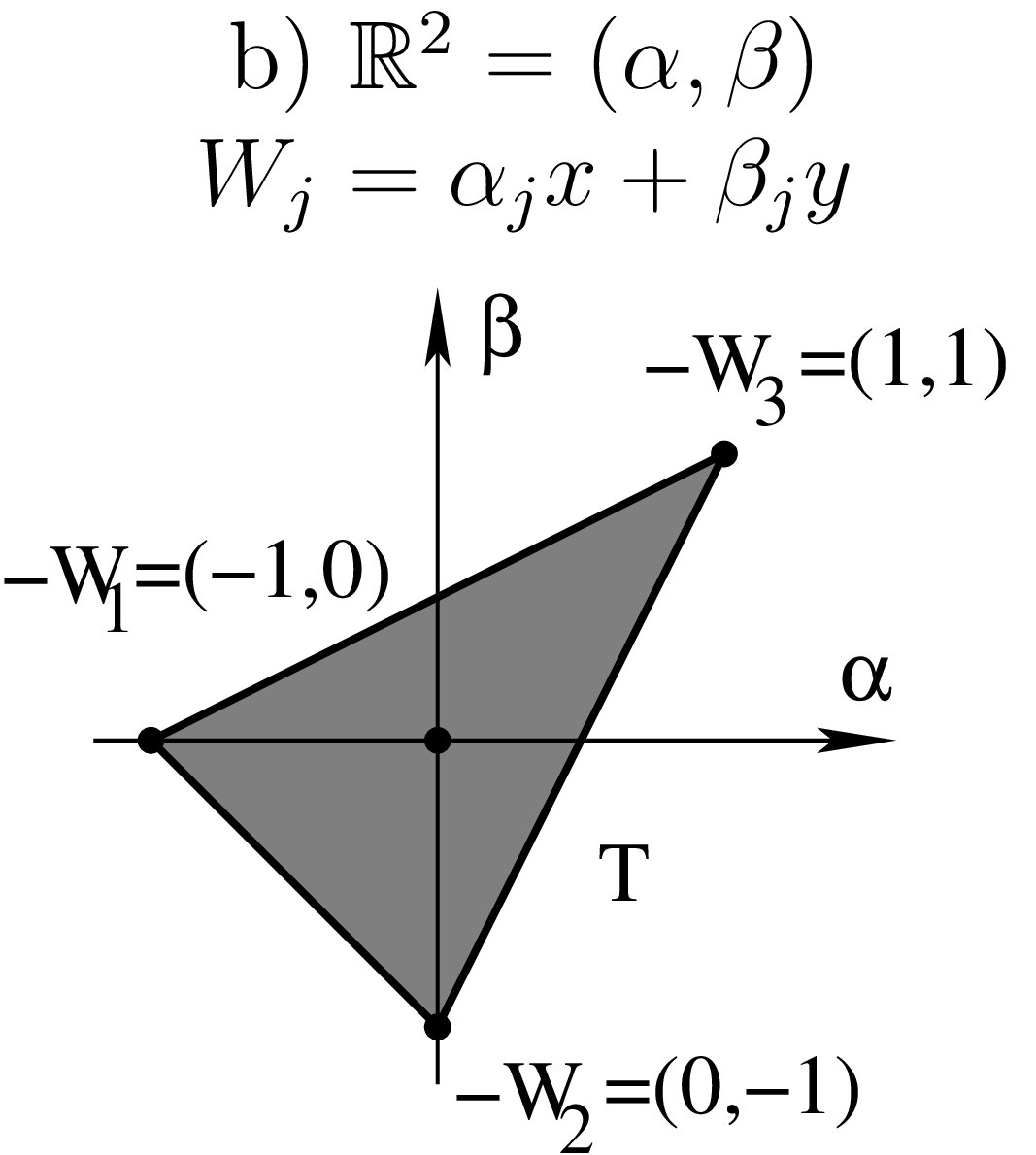}}
\end{center}
Here $e^{y}+e^{x}+e^{-y-x}=c$

Magnetic field is  decaying for $R\rightarrow
\infty$ except some selected  angles, it is a
Lump Type Field analogous to the KP ''Lump Potentials''. A linear
sum under the $1/2\Delta\log()$ reflects the Linearization of the
Burgers Hierarchy in the variable $c$\\
$$[B]=\int\int_{D^2_R}Bdxdy=-1/2R\oint_{S^1}I_{\{W\}}(\phi)d\phi+O(R^{-1}).$$
{\bf All points in $T_c$ define ground states in the
Hilbert Space $L_2({\mathbb R}^2)$. The boundary points  define the bottom of
continuous spectrum.}
Integral of square of eigenfunctions $c^{-1/2}$ is finite for $c$ inside $T$:
$\int_{{\mathbb R}^2}dxdy/c<\infty$. It can be easily checked for example in the example above $c=e^x+e^y+e^{-x-y}$,
where the integral integral can be easily investigated.

\pagebreak

{\bf The Periodic Problem.}

Let lattice in ${\mathbb R}^2$ be rectangular and $z=x+iy$. For every real periodic
 function $c$ we can define a whole family of ''possible'' meromorphic
Bloch functions\\
$$
 \psi''_{ext,\pm}=f(z)(\sqrt{c})^{\pm}e^{uz-\zeta(p)z}\frac{\sigma(z+p+R)}{\sigma(z+R)}
$$
where $f(z)$ is an arbitrary elliptic function.

 We have
$Q^+\psi''_{ext,-}=0$ for $L=L^+=QQ^+$.
 For anti-holomorphic functions we get  $Q\psi''_{ext,+}=0$
 for $L^-=Q^+Q$.\\
Let $c\neq 0$. We need  only nonsingular functions, so our manifold is
  $u\in {\mathbb C}P^1=\Gamma''$ with function $\psi''_+=e^{uz}\sqrt{c}$ (or $e^{u\bar{z}}\sqrt{c}$).

Let $c$ has an isotropic zero. We have larger family
of nonsingular (or ''weakly singular'') $\psi''_+$-Bloch functions
 because $\sqrt{c}/\sigma(z+R)$  became only weakly singular such that it may enter spectrum.
So the full manifold is $M^2={\mathbb C}P^1\times \Gamma$ where $\Gamma$ is an elliptic curve.
 If $c$ is a trigonometric polynomial, $1/c$ is not.
We have $c$ for $L^+$, so we have $c'=1/c$ for $L^-$. So we need to calculate $\psi'$ for
all real periodic smooth functions $c'$. \\
It will be done below after analysis of the genus 1.

\vspace{0.3cm}

{\bf 7. Solutions of genus 1, $\delta$-terms.}

\begin{center}
{\color{Black}Fig 3}
 \vspace{2mm}
  \mbox{\epsfysize=8cm\epsffile{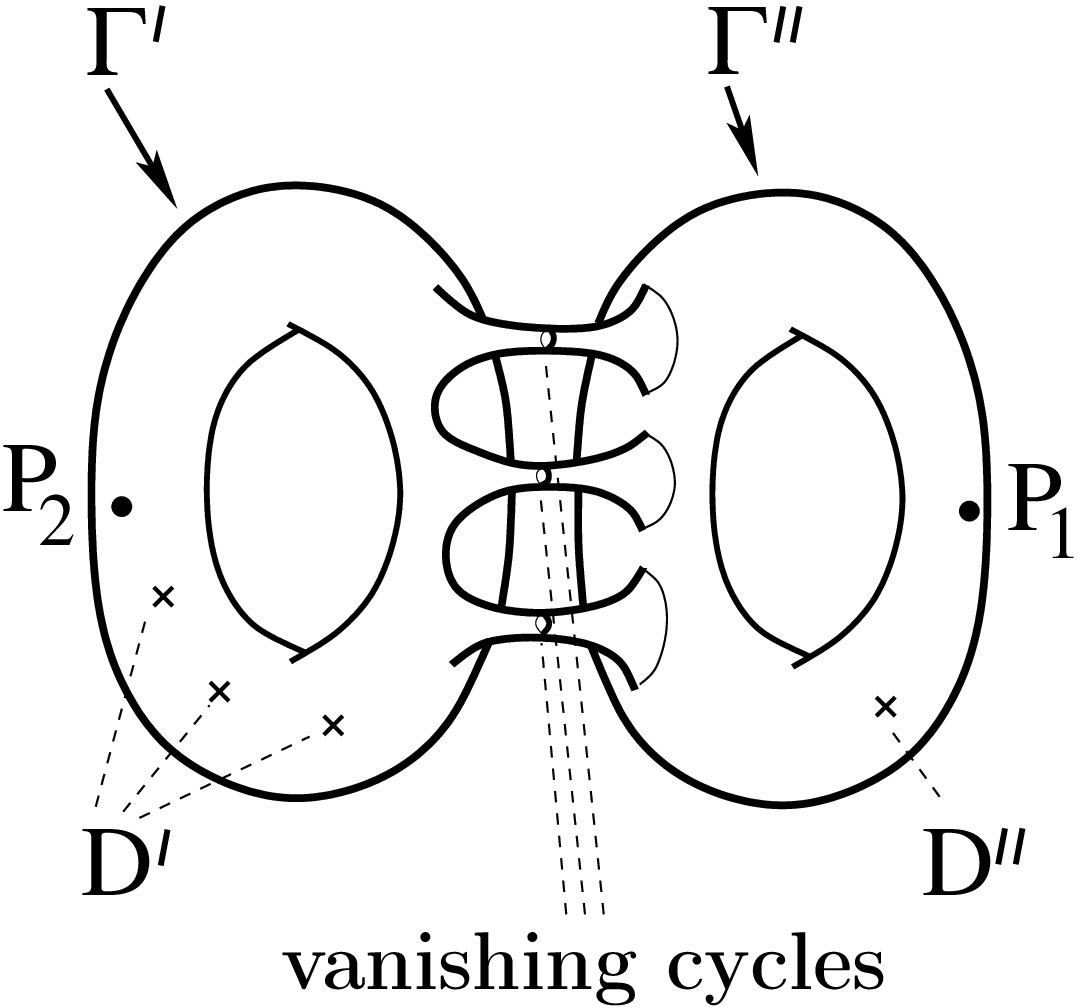}}\\
\end{center}

We take elliptic curve
$\Gamma'=\Gamma''={\mathbb C}/\Lambda$ with Euclidean local parameters $k,p$
(the point $0$ is ''infinity''), periods $2\omega\in {\mathbb R},2i\omega'\in i{\mathbb R}$ , $n$
intersection points $Q_0,...,Q_n\in \Gamma'$ and
$R_0,...,R_n\in\Gamma''$. Divisors $D'=(P_1,...,P_n),D''=P$ have
degree $n+1,1$ correspondingly. We have
$$
 \psi'=e^{-\bar{z}\zeta(k)}\frac{\prod_s\sigma(k-Q_s)}{\prod_l\sigma(k+P_l)}
 \times\left(\sum_j
 w_j\frac{\sigma(k+\bar{z}+\tilde{P}+\tilde{Q}-Q_j)}{\sigma(k-Q_j)}\right).
$$
Here $\tilde{P}=P_1+...+P_n,\tilde{Q}=Q_0+...+Q_n$, sum as in ${\mathbb C}$

$$
 \psi''=e^{-z\zeta(p)}\frac{\sigma(p+z+P)}{(\sigma(z+P)\sigma(p+P))},\
 \psi'(Q_s)=\psi''(R_s).
$$
All singularity of the quantity $c$
disappear after multiplication
$$
 \tilde{c}=c\sigma(\bar{z}+\tilde{Q}+\tilde{P})\sigma(z+P).$$
 Take
$n=1, Q_0=-Q_1,R_0=Q_1,R_1=Q_0$ and solution to the equation\\
$\omega\zeta(Q_0)=\eta_1Q_0$. We have $P=\tilde{Q}+\tilde{P}$ in this
case, so $-1/2\Delta |\sigma |^2=-2\pi \delta(z)$.

So  {\bf the Conclusion} based on the case $g=1$ is following:\\
The least singular real Data lead finally to magnetic field $\tilde{B}=-1/2\Delta\tilde{c}$
 which is periodic
nonsingular with magnetic flux equal to ONE QUANTUM UNIT. The original AG topologically trivial
magnetic field $B=1/2\Delta c$ extracted from our Inverse Spectral Data is always singular for
 $g=1$; it has total magnetic flux
equal to zero through the elementary cell and quantized $\delta$-singularity at the phone of
 nonsingular field $\tilde{B}$,
located in the point $P$.
 So this field corresponds to the ''Aharonov--Bohm''
(AB) situation.\\
For $g>1$ the number of quantized $\delta$-functions is
equal to $k>1$. {\bf We can get nonsingular AG operators only from genus $0$}.
Both pieces of the original Riemann surface $\Gamma=\Gamma''\bigcup\Gamma'$ are
 presented in the form of $k$-sheeted branching covering
over elliptic curve $\Gamma''\rightarrow \Gamma_0$ for $g>1$
as it was in the works of Krichever dedicated to the elliptic KP (see \cite{K2}).

{\bf The important Conclusion is following:}
Comparison with \cite{DN1} shows that the Quantized $\delta$-flux does not affect
 spectrum nearby of the zero level.

 The  complex Bloch--Floquet manifolds
 (consisting of nonsingular or weakly singular Bloch functions admitted by the spectral problem)
  for the level
 $\epsilon=0$ and genus $g=1$ is $$M=M^2$$ with function
$$
 \psi''_{ext,-}=\\\frac{1}{\sqrt{c}}\left(e^{uz}\times e^{-\zeta(p)z}\ \frac{\sigma(z+p+R)}{\sigma(z+R)}\right),
$$
  $$L^+\psi''_{ext}=L^+\psi'=0.$$
Let us point out that both functions $\psi',\psi''$ are originally  singular here $g=1$;
Our next step is  to reduce operator
 to the self-adjoint form $L$. For that we have to multiply both $\psi',\psi''$ by the
 factor $1/\sqrt{c}$. Let us start with $\psi''$. The function $\psi''_{ext}1/\sqrt{c}$
is ''nonsingular'', i.e. it is singular weakly enough
 for entering
   spectrum in $L_2({\mathbb R}^2)$ for the operator $L$ with the quantized Aharonov--Bohm term.
  After singular unitary gauge transformations $L\rightarrow \tilde{L}$ all singularity of
  this function disappear. So  it
    enters the smooth spectrum of the
   operator $\tilde{L}$
  with magnetic field $\tilde{B}$ where singularity is removed. Now about $\psi'$. The function
  $\psi'/c$ is nonsingular but the function $\psi'/\sqrt{c}$ is singular. After the singular
   unitary gauge transformation it maps into function $\tilde{\psi'}/\sqrt{c}$ with simple pole.
    It satisfies to the equation
$$
 \tilde{L}(\sqrt{\sigma/\bar{\sigma}}(\psi'/\sqrt{c}))=0.
$$
   {\bf So we conclude that the family $\tilde{\psi'}/\sqrt{c}$ of magnetic Bloch functions
    for the operator $\tilde{L}$
   with nonsingular topologically nontrivial magnetic field
   has a pole in the variables $(x,y)$.}\\ The  function $\psi'/c$ is nonsingular. It
   satisfies to the equation $L_0(\psi'/c)=0$
   in all nonsingular points
   where $L_0=\partial(\bar{\partial}+\Phi_z)$. At the same time we have $\Phi_z\sim 1/\bar{z}$
    near singularity, all other terms are smooth.
    So the function
    $f=(\bar{\partial}+\Phi_{z})(\psi'/c)$, $c=e^{\Phi}$,
   is antiholomorphic almost everywhere. However,  in the point $P$ where singularity
    is located, $f$ has a form $f\sim m(k)/\bar{z}$. Therefore  $\psi'/c\sim m(k)+\bar{z}O(1)$
     near this point. We conclude that\\ $L_0(\psi'/c)=m(k)\delta_P(x,y)$ because
     $\partial(1/\bar{z})=
     (const)\delta$.

{\bf Reconsider now the case $g=0$ comparing it with $g=1$.}\\
For $c\neq 0$ and $g=0$
 the Bloch manifold  is equal to the union $\Gamma''\bigcup \Gamma'$, and both are ${\mathbb C}P^1$;
  Let $c$ has an isolated zero
 (minimum) which is isotropic. Magnetic field became singular, with $\delta$-term.
 The extended instanton Bloch function $\psi''_{ext,+}$ became only weakly singular
 and capable to serve spectrum
 for the operator
 $L^+$. It depends  on the points of the complex 2-manifold $M^2$, and all possible values
 of the complex
 multipliers
$$
 M^2\rightarrow (\varkappa_1,\varkappa_2)\in {\mathbb C}^*\times {\mathbb C}^*
$$
  are presented in this family. The energy dispersion relation
 is degenerate here $\epsilon:M^2\rightarrow 0$.

 We have $M^2={\mathbb C}P^1\times \Gamma_0$
 where $\Gamma_0$ is an elliptic curve,
$$
 \psi''_{ext,+}=(const(u))e^{p\bar{z}-\zeta(u)\bar{z}}\sigma
 (\bar{z}+u)\sqrt{c}/\sigma(\bar{z}).
$$
Its singularity is in fact like $\sqrt{\bar{\sigma}/\sigma}$ because $c$ has a lattice of
double zeroes like $|\sigma|^2$. Make singular gauge transformation
 $$\psi''_{ext,+}\rightarrow \sqrt{\sigma/\bar{\sigma}}\psi''_{ext,+}$$
(here $\sigma$ depends on $\bar{z}$). This function
 transforms into  the nonsingular magnetic Bloch function
of the operator $L^P$ with topologically nontrivial magnetic field $\tilde{B}$. Magnetic
Bloch multipliers after transformation are equal to the original standard Bloch multipliers.

 Our Conclusion is that the periodic case $g=1$ gives  result similar to
 the special case of $g=0$ where $c$ has an isolated isotropic zero, interchanging sectors $\pm$
 (i.e. $z$ and $\bar{z}$).
 {\bf The higher number $k\geq 1$ of isotropic zeroes for $g=0$ leads to the
 ''higher rank'' family of weakly singular instanton type Bloch functions $M^{k+1}$}.
 Removing $\delta$-singularities
 by the singular gauge transformations we get smooth magnetic Bloch manifold $M^{k+1}$
 corresponding to the periodic magnetic field like in \cite{DN1}
  with
 higher flux.

\vspace{1cm}

{\bf 8. Extension of results to the infinite trigonometric series for $g=0$.}

\vspace{0.3cm}

 We know that the algebro-geometric case simply corresponds to the case
 of trigonometric polynomials. We take rectangular lattice in the plane $x,y$.
  Following relation is true
$$
 Q^+\psi'=M(k)\sqrt{c}e^{\bar{z}k}.
$$
  Now we choose normalization of $\psi'$ such that
  $M(k)=k$.

We use for that the formula
$$
 \psi'=k\sum_j \left[\frac{\kappa_je^{p_jz-k_j\bar{z}}}{k-k_j}\right]e^{k\bar{z}}.
$$
 We are dealing here with the
 new normalization: here $k_j$ are simply the lattice points.\\
 As a corollary we have $$\sum_j\kappa_je^{p_jz-k_j\bar{z}}=c$$\\
Apply this result to the infinite series $c\rightarrow c'=1/c$.
It gives us a
function $\psi'_{-}$ for the second component  $L^-$ of the Pauli operator.\\

\vspace{1cm}

{\bf 9.The Boundary Problems}.

\vspace{1cm}

{\bf Problem}: The component $\Gamma'$ of the Bloch--Floquet manifold does not affect the
ordinary spectrum
in the Hilbert space of functions in the whole plane ${\mathbb R}^2$.
 Can we use it for solving physically meaningful (i.e. self-adjoint) boundary problems?\\

 Let us remind following: For  one-dimensional  self-adjoint periodic Schr\"o\-dinger operators
 $L=-\partial_x^2+u,\ u(x+T)=u(x)$, the 2-sheeted Riemann surface $\Gamma$ of the Bloch--Floquet
 function
 $L\psi=\lambda\psi,\ T^*\psi=\psi(x+T)=\varkappa\psi$,
 is such that all branching points $a_s, s\geq 0,$  are real. They are  the boundaries
  separating the
  spectral zones
 from the forbidden zones (gaps) at the real line. In order to reconstruct potential,
 we need to know also the position of
  poles of  $\psi(\lambda,x,x_0)$ for $\lambda\in\Gamma$ (normalized by condition
  $\psi(x_0)=1$). They
   are the points on
  the Riemann surface $\gamma_j\in \Gamma$ forming together a ''Divisor $D$''. There is
 exactly one pole inside of every
 finite gap for the smooth potentials:
$$
  \gamma_j(x_0)\in [a_{2j-1},a_{2j}].
$$
 The zeroes of $\psi$  are located exactly in the points
 $\gamma_j(x)$. Therefore our Bloch--Floquet function $f(x^*)=\psi(\gamma_j(x),x^*)$ satisfies
  to the Dirichlet
 boundary problem at the interval $x^*\in [x,x+T]$ where $T$ is a period because $f(x)=f(x+T)=0$.

 Another interpretation follows from the fact that the same function $f(x^*)$ decreases to
 the right or to the left from the zero $f|_{x^*=x}=0$ corresponding to
 the sheet of Riemann surface
 where the pole $\gamma(x_0)$ is actually located. So the function $f(x*)$
  belongs to the discrete spectrum on the
  corresponding half-line $x^*\in [x,+\infty]$ or $x^*\in [-\infty,x]$.

{\bf Question: Can we find similar realization for the  Bloch functions $\psi',\psi''$
for the periodic smooth topologically trivial magnetic field $B=1/2\Delta \ln c$ where $c$
is real and nonzero?}

The classical John von Neumann theory of self-adjoint boundary conditions
(as it was observed by I. Gelfand in his discussion with  S. Novikov in the begining of 1970s) follows in fact to
 the
scheme  quite similar to the
  Novikov's symplectic idea to construct
 the ''Hermitian K-theory'' over rings
  with involution: Self-adjoint extensions of symmetric operators correspond to Lagrangian
  subspaces in the
  Hamiltonian module over the ring of functions. Let $\gamma\subset T^2$ be a contour in the
  torus presenting the boundary of our
   domain (or its connected component).

   Consider first the ''scalar'' boundary conditions
   for the operator
   $L=L^+=QQ^+$ or $L=L^-=Q^+Q$ separately, not mixing different components. We reduce
    the integral to the boundary
   $$\int\int_{D}[(L\psi)\bar{\phi}-\psi(\bar{L}\bar{\phi})]d^2x=\oint_{\gamma}[\psi_1
   \bar{\phi_2}-\psi_2\bar{\phi_1}]dt$$
where $\psi_1(t)=\nabla_n\psi |_{\gamma}=(\partial_n+A_n)\psi |_{\gamma}$ and
$\psi_2=\psi|_{\gamma}$. Here $n$ means the  vector  normal to the boundary component
 (external). To make  self-adjoint extension means to select a ''Lagrangian Subspace''
 $(\psi_1(t),\psi_2(t))\in\Lambda$
 for the  skew Hermitian boundary form such that
 $$\oint_{\gamma}[\psi_1\bar{\psi'_2}-\psi_2\bar{\psi'_1}]dt=0$$
for every pair of elements in the subspace $\Lambda$. There are
 following types of boundary conditions.

{\bf The ''Ultralocal'' boundary conditions such that no derivatives along the directions tangent
 to the boundary are involved}. They are elliptic.

1. The Dirichlet b.c $\psi_2=\psi |_{\gamma}=0$.\\
2. The Neuman b.c. $\psi_1=\nabla_n \psi |_{\gamma}=0$.\\
3. The Leontovich b.c. $\alpha(t)\psi_1+\beta(t)\psi_2=0$ with $(\alpha,\beta)$ real
 and nonzero for all $t\in\gamma$. Only ratio $(\alpha(t)/\beta(t))^{\pm 1}$ is invariant here.
 This b.c. has a topological charge (the degree of map $\gamma=S^1\rightarrow S^1={\mathbb R}P^1$).

{\bf The General ''Local'' b.c.}. They are nonelliptic.

 Let a pair of scalar differential operators on the circle $U,V$ be given such that one of them
 is invertible (for example, equal to $1$). Our b.c. is $U\psi_1=V\psi_2$. This b.c.
 is self-adjoint if $(UV^+)^+=UV^+=VU^+$. In the special case then one of them is equal to 1,
 another one is an arbitrary self-adjoint differential operator on the boundary. \\

  An interesting special case we have if $U=1$, and $V$ is a first order operator
 $V=i\alpha \partial_t+v(t),\alpha,v(t)\in {\mathbb R}$.
 The b.c. of the form $\psi_1=\nabla_n\psi=\pm i\psi_t+v(t)\psi$ we call the {\bf ''$\partial$-bar boundary condition''
 or ''$\partial$-boundary condition''}.
  The reason for that is following: we can write this b.c. (for the sign $-$) in the form
 $$[Q^+\psi-e^{i\theta(t)}v(t)\psi]_{\gamma}=0$$
 where $Q^+=-(\partial_{\bar{z}}-
 A_{\bar{z}})$ and $A_1=i\Phi_y,A_2=-i\Phi_x, A_{\bar{z}}=-\Phi_{\bar{z}}, \theta(t)$
 is an angle rotating the coordinate
 frame
 $x,y$ to the Frenet frame $n,\partial_t$. The $\partial$-bar boundary condition is self-adjoint but not ultralocal.
  It is
 nonelliptic. We could not find any traces of it in the literature. However, it is very closed
 to the elliptic boundary condition and self-adjoint. One can imagine that such boundary condition might appear in
  the problems of physics. Replacing the operator $V$ by $V'=+i\partial_t+v(t)$ we obtain similar
  {\bf ''$\partial$-boundary condition''} where $Q^+$ is replaced by $Q=\partial-\Phi_z$. It has similar properties.

\vspace{0.5cm}

{\bf  There are also  General Nonlocal Self-Adjoint boundary conditions}.
We do not consider here nonlocal boundary conditions.

\vspace{1cm}

{\bf  Boundary Conditions for  Pauli operators mixing components.}

The operator $L^P$ acts on vector-functions $\Psi=(\psi^+,\psi^-)$. After calculations
we obtain  $$<L^P\Psi,\Phi>-<\Psi,L^P\Phi>=\oint_{\gamma}\{[\psi_1^+\bar{\phi}_2^+-\psi^+_2
\bar{\phi}_1^+]+[\psi^-_1\bar{\phi}^-_2-\psi^-_2\bar{\phi}^-_1]\}dt$$
where
$$
 \psi^{\pm}_1=\nabla_n\psi^{\pm} |_{\gamma},\quad \psi^{\pm}_2=\psi^{\pm} |_{\gamma}.
$$

{\bf The Ultralocal boundary conditions.} They are elliptic.

Our skew Hermitian form has signature $2,2$ in every point of boundary $t\in\gamma$.
The family of all Lagrangian subspaces is isomorphic to the group $U_2$ for every $t\in\gamma$.
So all collection of ultralocal self-adjoint b.c. can be classified by the smooth maps
$S^1\rightarrow U_2$ with topological charge from the group $\pi_1(U_2)={\mathbb Z}$.
Lagrangian subspaces in every point $t\in\gamma$ may have 3 types:\\
1. Subspaces of the form $$\psi_1^{\alpha}=\sum_{\beta}R^{\alpha,\beta}\psi^{\beta}_2$$
where matrix $R$ is Hermitian and $\alpha,\beta=\pm$.\\
2. Similar subspaces with numbers 1 and 2 interchanged $\psi_2=R\psi_1$.\\
The intersection of the types 1 and 2 is exactly a subclass such that Hermitian matrix $R$ is
 nondegenerate.\\ The types 1 and 2 can be unified by the form $U\psi_1=V\psi_2$ where one of
  matrices $U,V$ is invertible and $UV^+=VU^+$ is Hermitian.\\
3. Subspaces of the form $a\psi_1^++b\psi_1^-=0, \ c\psi_2^++d\psi_2^-=0$.
Lagrangian property implies $a\bar{c}+b\bar{d}=0$. We can normalize these equations
reducing them to the form $|a|^2+|b|^2=1,\
|c|^2+|d|^2=1$. We are coming to the arbitrary matrix from the group $SU_2$ numerating such
special Lagrangian subspaces (or to the arbitrary unit vector $(a,b)\in {\mathbb C}^2$ determining
 the whole matrix).

\vspace{0.2cm}

{\bf The General Local boundary condition mixing components}.\\
 We can write a natural differential analog of the scalar types above in the form
$$U\psi_1=V\psi_2,\quad \psi_1=\nabla_n\psi,\quad \psi_2=\psi$$ where one of two  operators $U,V$
 on the circle is invertible.
Our requirement is $UV^+=VU^+$.\\
The most interesting for us are the boundary conditions where derivatives of both components of $\Psi$ in
the boundary points are involved
$$\nabla_n\psi^+=A^+\psi^++b\nabla_n\psi^-,$$
$$-\psi^-=c\psi^++d\nabla\psi^-$$
where $A^{+}=\alpha^{+}i\partial_{\tau}+a$ are the first order operators along
 the boundary. Here the functions $\alpha^{+}(t), a(t),d(t)\in {\mathbb R}, b(t),c(t)$
  are smooth and $c=\bar{b}$.

This boundary condition is nonelliptic if order of operator $A^{+}$ is more than zero. The ultralocal case
corresponds to the zero order $\alpha^{+}=0$.

\pagebreak

{\bf 10.Bloch functions outside of the spectrum as solutions of the
boundary problems.}

\vspace{0.3cm}

We discuss the following Problem.\\
 Let $\Psi=(\psi^+,\psi^-)$ be a formal zero mode solution to the Pauli operator
 $L^P\Psi=0$.
For which contours it  satisfies to some self-adjoint boundary condition?

We are going to consider below
 this Problem both for mixing and nonmixing cases.

 {\bf Our intention is to show that for some
 special boundary conditions these contours can be described as a trajectories of
 dynamical system (foliation)
 in the 2-torus $T^2$ or in the covering space ${\mathbb R}^2$.} Consider first the Ultralocal case.

\begin{thm} Let $d=0$ and $\alpha^+=0$. The coefficients $a,c,b$ are such that $a\in {\mathbb R}$
 and $c=\bar{b}$ if and only if the boundary curve $\gamma=\partial D$ is a leaf of foliation
 $\omega=0$ where $$\omega=(|\psi^+|^2+|\psi^-|^2)d\Phi+|\psi^+|^2*d\theta^+
 +|\psi^-|^2*d\theta^-$$
Here $\psi^{\pm}=|\psi^{\pm}|\exp\{i\theta^{\pm}\}$ and $*d\Phi$ is a vector potential
 of magnetic field,
star $*$ is a standard duality operator of 1-forms on the complex plane.

\end{thm}

The case $b=0$ leads to the nonmixing b.c.. We will discuss this case below in details.
For
the $\partial$-bar problem we have  $\alpha^+=-1$. Assuming as above that $d=0$ we are coming
to the similar dynamical system.
It is investigated below in details for the nonmixing case $b=c=0$.

The proof of this theorem is not complicated. It can be obtained by the elementary manipulation.

{\bf The case which is most interesting for us is $\psi^+=\psi'$ and $\psi^-=\lambda\psi''$ where
 $\lambda$
is a parameter, and $\psi''=Q^+\psi'$.} Here $Q^+$ is the operator which is treated in
 quantum theory as a supersymmetry for $L^P$
(see above). Our dynamical system depends on $\lambda$.
For $\lambda=0$ we have a Nonmixing Case.

 This pair defines a General
Bloch Solution of the zero energy level
to the Physical Pauli Operator with some (nonunitary) multiplier like in the one-dimensional case where
we had Bloch solutions in the forbidden bands as eigenfunctions of the self-adjoint boundary problem.

\vspace{0.3cm}

Let us make some very general remarks concerning the
 {\bf Nonmixed Boundary Problems:}

Let a contour $\gamma\subset T^2$ be given with following Domain $D=T^2\setminus\gamma$.
Find solution to the boundary problem 1 or 2 such that $L^+\psi=0$ in $D$ and

{\bf The Ultralocal Problems}. Let $\alpha(t)\nabla_n\psi+\beta(t)\psi=0$ at the contour $\gamma$.
Consider first the Problem 1 with  {\bf contours homotopic to zero in $T^2$} (see Fig 4 a),
$\partial D=\gamma$. Solution of the  ''Leontovich type'' boundary problem $\psi$
satisfying to this relation
should be found from the integral equation
$$
 \psi =\int\int_{{\mathbb R}^2}[p(k)\psi'+q(k)\psi'']d^2k.
$$

 For the Problem 2 with {\bf contours $\gamma$ nonhomotopic to zero in $T^2$} we have
 $\partial D=\gamma
\bigcup \gamma'$ (see Fig 4 c). There is a natural
${\mathbb Z}$-covering $\hat{D}\rightarrow D$ which is a strip
in ${\mathbb R}^2$ with average direction $g=[\gamma]\in H_1(T^2,{\mathbb Z})={\mathbb Z}^2$
 (see  Fig 4 b). The element
$g\in {\mathbb Z}^2\subset {\mathbb C}={\mathbb R}^2$
acts freely  $g:\hat{D}\rightarrow\hat{D}$ as a shift by the complex number $g:z\rightarrow z+g,
\bar{z}\rightarrow\bar{z}+\bar{g}$.
 By definition, the components of boundary
$$
 \partial \hat{D}=\hat{\gamma}\bigcup
 \hat{\gamma'}
$$
are coverings over $\gamma,\gamma'$.
$g$ can be viewed as a vector of lattice ${\mathbb Z}^2$
in ${\mathbb R}^2$. Another generator $g'\in {\mathbb Z}^2$ complementary to $g$ maps exactly $g':\gamma\rightarrow
\gamma'$. We are saying that $\gamma'(\hat{\gamma'})$ {\bf is located to the right from $\gamma
(\hat{\gamma})$}
 if $|\varkappa(g')|<1$.
Let $c\neq 0$. All functions $\psi'(k),\psi''(k)$ with $k$ orthogonal to $g$
have unitary Bloch multipliers $|\varkappa(g)|=1$ along the shift $g$. We have 4 domains
$D_+,D_-,\hat{D}$
in ${\mathbb R}^2$ and $D$ in $T^2$ (see Fig 4):\\
$$
 D_+\bigcup D_-={\mathbb R}^2,\ D_+\bigcap D_-=\hat{\gamma},\ g'(D_-)\subset D_-,\
 g'(D_+)\bigcap D_-=\hat{D}.
$$
We require $\alpha(t)\nabla_n\psi+\beta(t)\psi=0$ on the contours $\gamma,\gamma'$
{\bf but allow to have
different pairs of coefficients $(\alpha,\beta),(\alpha'\beta')$ for $\gamma$ and $\gamma'$ for the
domain $\hat{D}$ such that
 both of them enter the boundary}.
The function $\psi$  should be constructed essentially from $\psi'$ but we can add also
 $\psi''$ if necessary: The most general possibility
is that $\psi$ is given by the integral along the $k$-axis orthogonal to $g$:
$$
 (I):  \psi=\int_{k\geq 0}[\psi'(k,x,y)p(k)+\psi''(k,x,y)q(k)]dk
$$
for the half-plane $D_+\subset {\mathbb R}^2$\\
$$
 (II):  \psi=\int_{k\leq 0}[\psi'(k,x,y)p(k)+\psi''(k,x,y)q(k)]dk
$$
for the half-plane $D_-\subset {\mathbb R}^2$\\
$$
 (III): \psi=\int_{k\in {\mathbb R}}[\psi'(k,x,y)p(k)+\psi''(k,x,y)q(k)]dk
$$
for  the strip $\hat{D}\subset {\mathbb R}^2$
$$
 (IV)_{\varkappa}: \psi_{\varkappa}=\sum_{m\in {\mathbb Z}}[\psi'(k_m,x,y)p_m(\kappa)+\psi''(k_m,x,y)]q_m(\kappa)
$$
 for $D\subset T^2$, where
corresponding $k_m$ are such that all $\psi'(k_m),\psi''(k_m)$ have the same fixed  unitary
multiplier $\varkappa=e^{k_m\bar{g}}$ (i.e. $k_m=k_0+2\pi im/\bar{g}$ where $m\in {\mathbb Z}$).

{\bf The local ``$\partial$-bar''  (or ``$\partial$-'') type Boundary Problems}.

As above in the Ultralocal case we have two types of contours :\\
Problem 1 with contours $\gamma$
homotopic to zero in $T^2$
where $\gamma=\partial D$. \\
 Problem 2 with contours  $\gamma\subset  T^2$ non-homotopic to zero in the torus; here
$\partial D=\gamma\bigcup\gamma'$.\\

For the second case (most interesting for us) we consider the same type of domains
$D_+,D_-,\hat{D},D$ with boundaries non-homotopic to
zero in $T^2$  and their coverings in ${\mathbb R}^2$ as above (see Fig 4).

\noindent
\parbox[s][8cm]{4.5cm}{\begin{center}\epsfxsize=4.2cm\epsffile{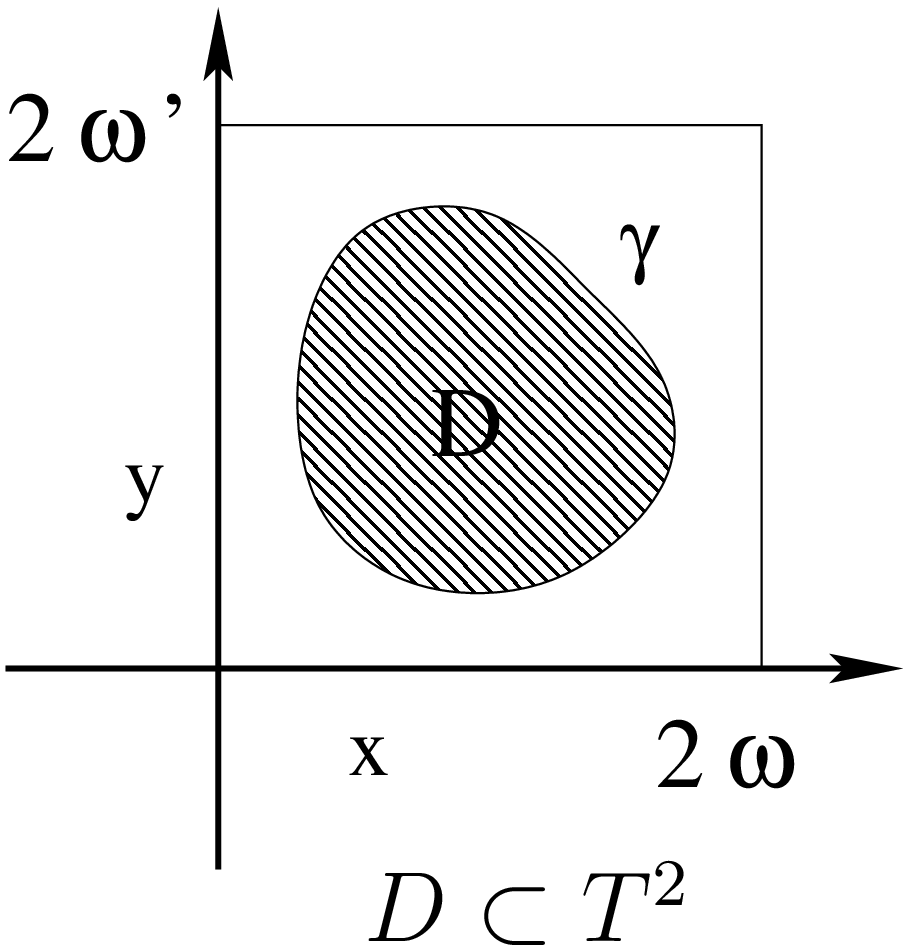}

\vspace{2cm}

Fig 4a \\ $\gamma$ is homotopic to 0 \end{center}}
\parbox[s][8cm]{9cm}{\begin{center}\epsfxsize=8cm\epsffile{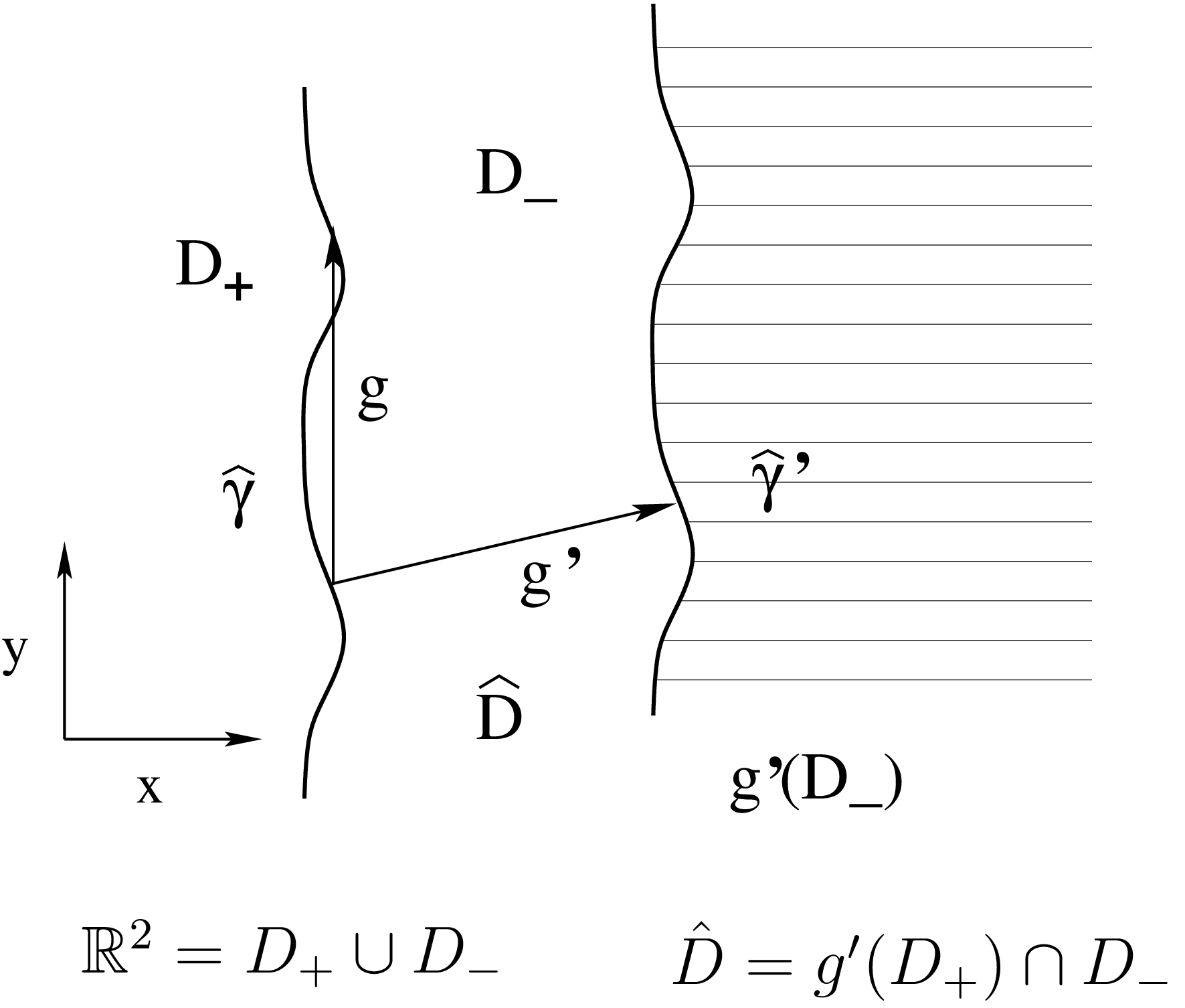} \\
 Fig 4b\\ $\gamma$ is not homotopic to 0 \end{center}}

\vspace{0.3cm}

Fig 4a and Fig 4b are associated with Example 2 below. Here $g$ is parallel to the average direction of
$\hat{\gamma}$

\vspace{1cm}

\noindent
\parbox[s][8cm]{6.5cm}{\begin{center}\epsfxsize=6cm\epsffile{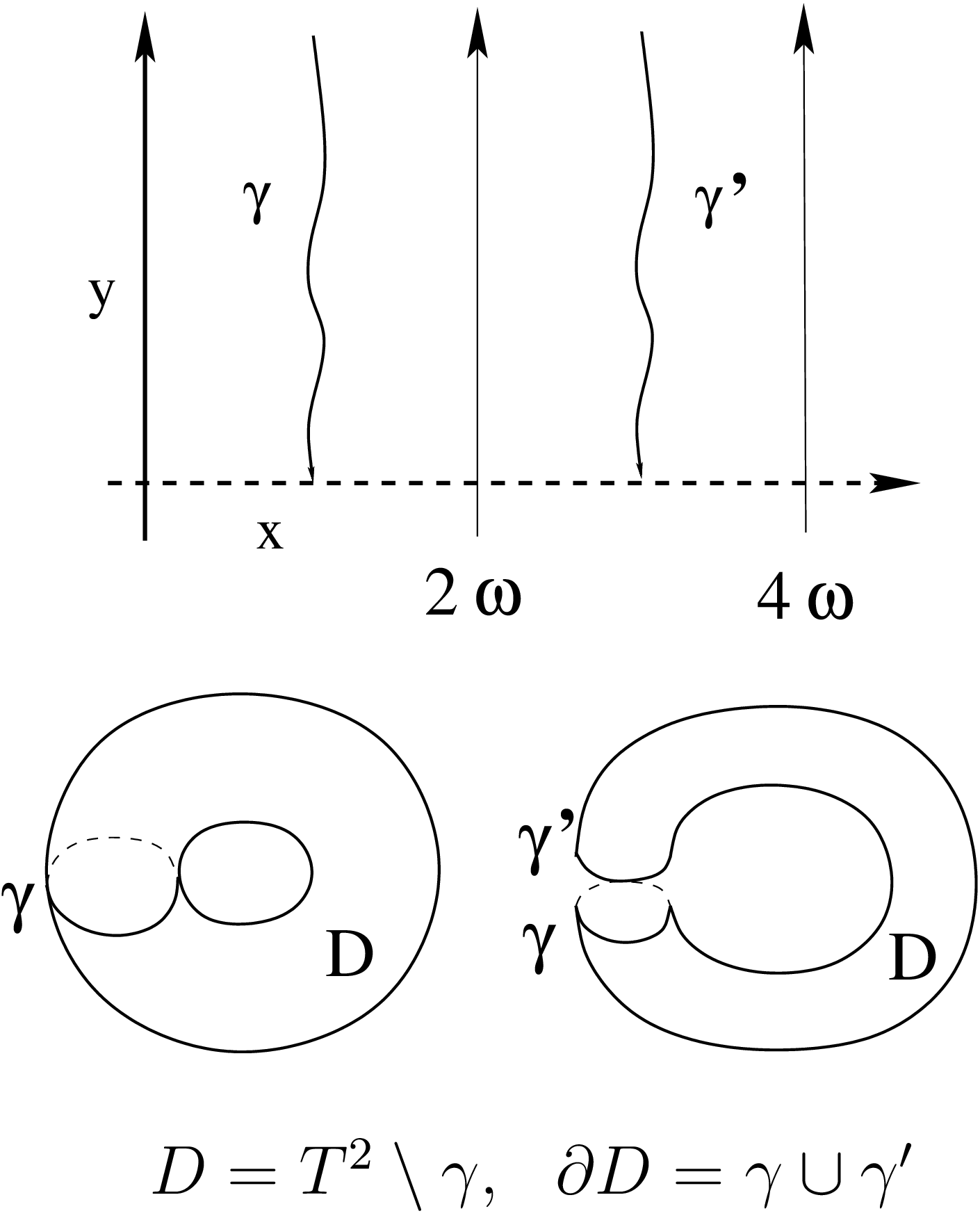}\\
 Fig 4c\end{center}} \hspace{1cm}
\parbox[s][8cm]{6.5cm}{\begin{center}\epsfxsize=6cm\epsffile{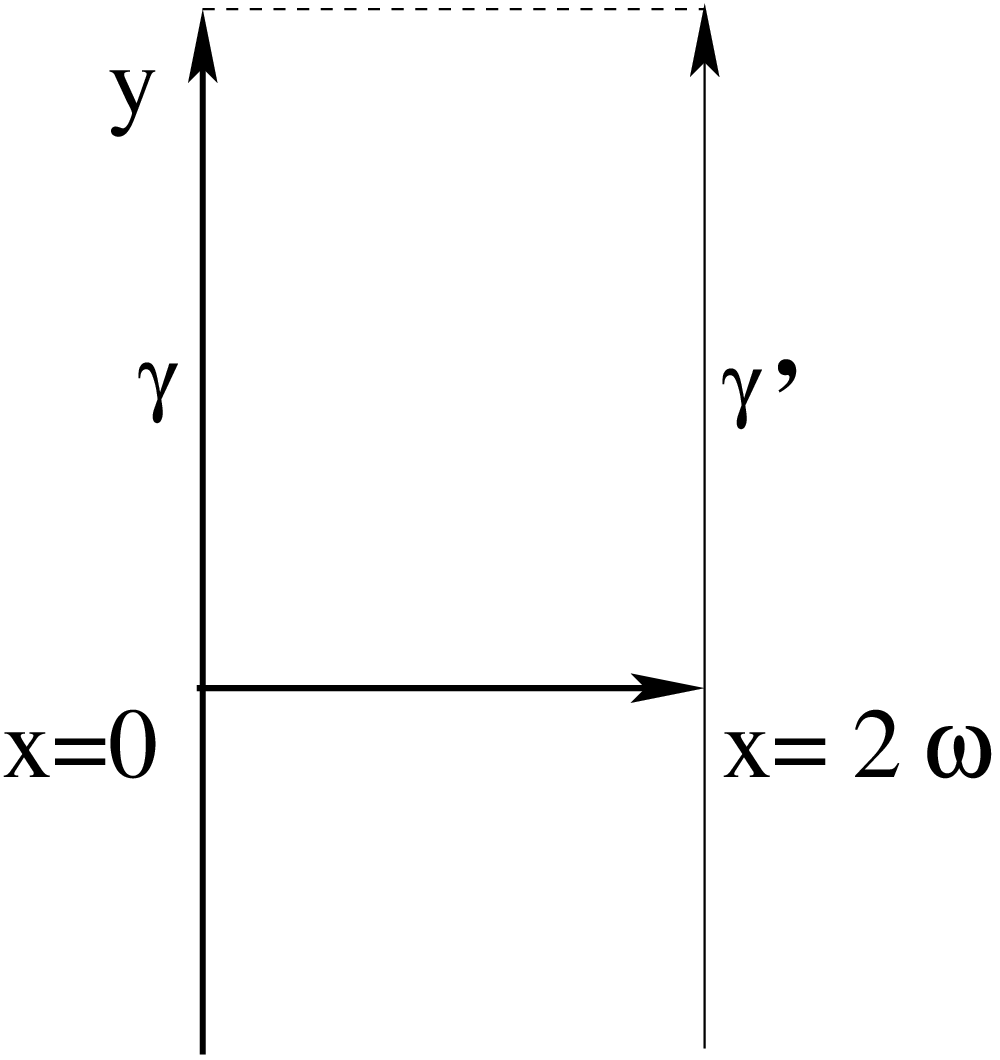}

\vspace{1cm}

 Fig 4d\end{center}}

\vspace{0.3cm}

Fig 4c and Fig 4d are associated with Example 1. The contour $\hat{\gamma}$ is simply $x=0,
 -\infty <y<\infty$ in Fig 4c.

\vspace{1cm}

 Our b.c. relation along the boundary
is (with possibly different coefficients in the different components of boundary)
$$Q^{\pm}\psi=v(t)\psi e^{i\theta(t)}$$ or
$$\nabla_n\psi=(\pm i\partial_t+u(t))\psi$$ for real $u,v$.
We can try to find solution $\psi$ using the integral equations as above.

\vspace{1cm}

We already discussed above the following question: {\bf which general restrictions should be satisfied for
  the class
 of functions $\psi$ solving the  Leontovich type  problem in the magnetic field
 with vector-potential $A=-i*d\Phi=i(-\Phi_ydx+\Phi_xdy)$?}

We consider now only nonmixing case.
 Assuming that $\psi=\rho(x,y) e^{i\theta(x,y)}$ we can see that the ratio
  $\nabla_n\psi/\psi$
 should be real. Here $n$ is a normal vector to the boundary contour $\gamma\subset\partial D$.
We are coming to the following statement (a partial case of the theorem above):

\begin{lem}The boundary contour $\gamma$ for the self-adjoint scalar operator $L\psi=\lambda\psi$
with any Leontovich type b.c.
should be a leaf of foliation given by the equation
$*(d\theta+A/i)=\Omega=0$ or $(\theta_y+\Phi_x)dx+(-\theta_x+\Phi_y)dy=0$ if the phase
 function $\theta(x,y)$ is known. Vice versa, if the contour $\gamma$ is known, the
 restriction $*d\theta=-d\Phi$ along the contour should be satisfied.
\end{lem}

\vspace{1cm}

{\bf The Special Contours where solution to the Nonmixing B.C. is a Bloch function $\psi'$.}

\vspace{0.5cm}

{\bf Example 1: $\psi'$ as a solution to the Leontovich Boundary Problem for the
 Special Contours of the first kind.}

Let $g=0$. We construct solution with $2n+1$ crossing points $p_j,k_j$
where
$$
 p_j=a+ib_j,\quad p_{n+j}=-p_j,\quad k_{n+j}=-k_j.
$$
We have constants (nothing to do with Bloch  multipliers)
 $\kappa_{n+j}=\kappa_j$
and all parameters $a,b_j,\kappa_j$ are real. Let $\sum_{j=1}^n\kappa_j=0$. We have
$$c=\kappa_0+2\sum_{j=1}^n\kappa_j\cos(2(b_jx+ay)),$$

$$
 \frac{\psi'}{k}=e^{k\bar{z}}\left(\frac{\kappa_0}{k}+\sum_{j=1}^n\left(\kappa_j
 \frac{e^{p_jz-k_j\bar{z}}}{k-k_j}
 +\kappa_{n+j}\frac{e^{p_{n+j}z-k_{n+j}\bar{z}}}{k-k_{n+j}}\right)\right).
$$
For $x=0$ we have $c=\kappa_0=const$. We call such contours {\bf  Special,
 of the first kind}.
 We take now contour $\gamma:x=0$ or $c=const$ with boundary problem
$$
 \psi_n=\alpha(x)\psi.
$$
 Let us point out here that for such contours we have $\nabla_n=\partial_n$ because $d\Phi=0$
 at the contour $\Phi=\ln c=const$ and $A=(i\Phi_y,-i\Phi_x)$.
 We need to satisfy condition $\psi'_x/\psi\in {\mathbb R}$ for $x=0$. Presenting $\psi'$ in the form
 $\psi'=e^{k\bar{z}}(f+ig)$ we see that $\psi'_x/\psi'|_{x=0}\in {\mathbb R}$ follows
 for the real $k\in {\mathbb R}$ from the requirements
 $g|_{x=0}=g_x|_{x=0}=0$. After elementary calculations we see that $g|_{x=0}=A_1\cos(ay)+A_2\sin(ay)$
 and $g_x|_{x=0}=B_1\cos(ay)+B_2\sin(ay)$ where the constants $A_q,B_q$ are linearly expressed
 through the constants $\kappa_j$ above. Our boundary condition is equivalent
  to 4 homogeneous linear equations
 $A_1=A_2=B_1=B_2=0$. So for $n>5$ we can find solution for every fixed values of constants
 $a,b_j,k\in {\mathbb R}$. The real values of $k$ are orthogonal to the direction
  of the contour $x=0$. So we proved following

 \begin{lem} For every set of exponents with  odd number $2n+1$ of crossing points and $n>5$,
 there exists a divisor (i.e. the set of constants $\kappa_j$) such that corresponding
   Bloch--Floquet function $\psi'(k,x,y)$ with real $k\in {\mathbb R}$
satisfies to the Leontovich b.c. $\psi'_x/\psi'\in {\mathbb R}$ for the contour $\gamma: x=0$.
 The Bloch multiplier along the direction of  contour  is unimodular
 $\kappa=e^{-iky}$.
 \end{lem}

  So our problem is self-adjoint. Corresponding function $\psi'(k,x,y)$
 serves as a solution to the Leontovich boundary problem in the domain
  $\hat{D}$: $ [0<x<2\omega ]$, and in one of two
  half-planes
 $D_+$: $[0<x]$ or $D_-$: $[x<0]$
 depending on sign of the corresponding value of $k$. It serves also for the spectrum in
  the domain $D$ where
 $x\in [0,2\omega]$ in the torus $T^2$ (i.e. periodic in the variable $y$), see Fig 4c and 4d.

\vspace{1cm}

Fig 5 below is associated with the next Example 2 dedicated to $\partial$-bar boundary conditions. It
shows situation listed in the final conclusions as a ``maximal'' boundary contour
 homotopic to zero
such that the area inside coincides with the whole area of the torus. In particular, the magnetic
 flux through this domain is equal to zero
according to our assumptions on the class of magnetic fields under consideration. In this case a
 boundary cycle consists
 of separatrices (see Fig 5a and Fig 5b ):
\begin{center}
{\color{Black}Fig 5}\\
 \vspace{2mm}
  \mbox{\epsfxsize=14cm\epsffile{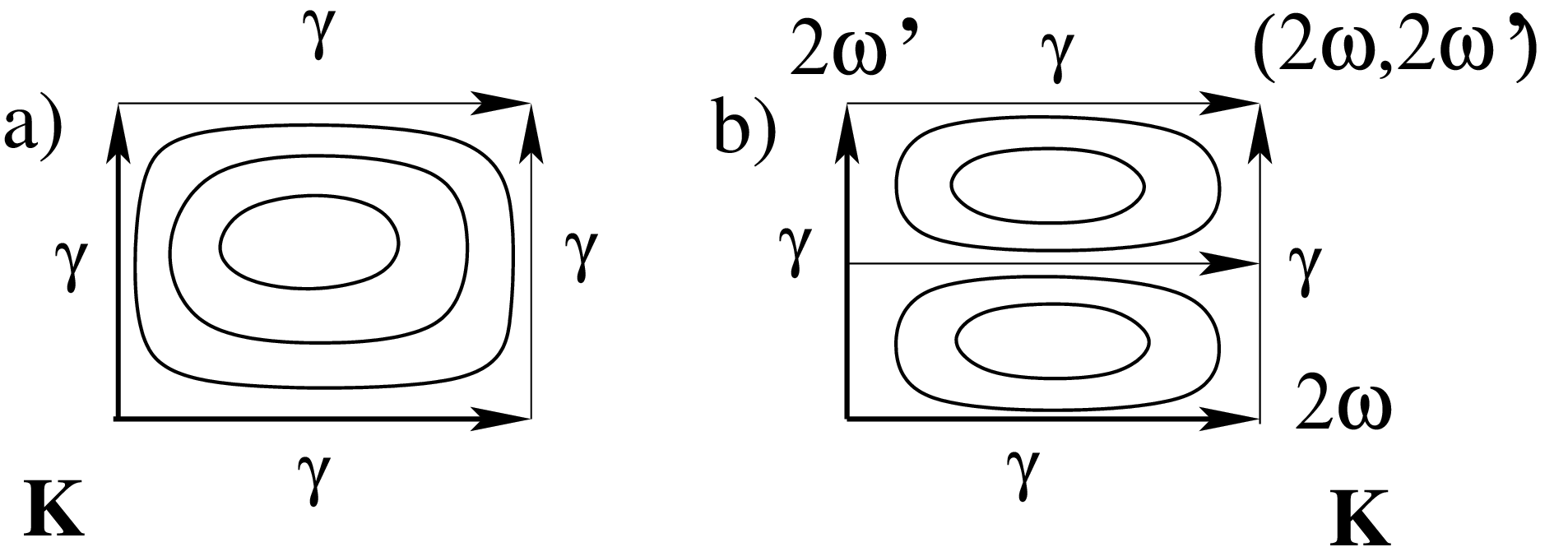}}\\
\end{center}

 \vspace{1cm}

 {\bf Example 2: $\psi'$ as a solution to the ``$\partial$-bar'' boundary problem for the special
 contours of the second kind.}

We write $\partial$-bar problem in the form $Q^+\psi=u(t)e^{i\theta(t)}\psi$ where $u\in {\mathbb R}$ and
$\theta(t)$ is an angle from the frame
$x,y$ to the Frenet frame $n,\partial_{\tau}$ along the boundary curve $\gamma$.
Let $\psi=\psi'(k,x,y)/\sqrt{c}$. We have $Q^+\psi/\psi=e^{i\theta_k(x,y)}u_k(x,y)$
 where $u_k\in {\mathbb R}$. So our contour $\gamma$ should be tangent
 to vector (director) field $\partial_{\tau}$ in the torus.
 This field depends on parameter $k\in {\mathbb C}$.
 It is  obtained from the direction $\partial_x$ by the rotation $e^{i\theta_k(x,y)}$
 in the point
$x,y$ or $z=x+iy$. This vector (director) field can be described by the zeroes of one-form
$$
 {\rm Re}[(Q^+\psi/\psi)d\bar{z}]=0.
$$
Let
$$
 c=1+\sum_j^na_je^{l_j\bar{z}-\bar{l}_jz}+\bar{a}_je^{-l_j\bar{z}+\bar{l}_jz}
$$
and in the new normalization where poles are located in the crossing points
$$
 \frac{\psi'}{k}=\left(\frac{1}{k}+\sum_j \left(a_j\frac{e^{l_j\bar{z}-\bar{l}_jz}}{k+l_j}+\bar{a}_j\frac{e^{-l_j\bar{z}
+\bar{l}_jz}}{k-l_j}\right)\right)e^{k\bar{z}}.
$$
 Now we make a gauge transformation to the self-adjoint form  $L=QQ^+$, and
 $\psi=\psi'/\sqrt{c}$. We have $Q^+\psi/\psi=-\bar\partial\psi'/\psi'$.
But $\bar\partial(\psi'/k)=2ce^{k\bar{z}}$ for the newly normalized $\psi'$.
 We obtain result
$Q^+\psi/\psi=-2cke^{k\bar{z}}/\psi'$.

 Therefore  we described  the set of contours for which $\partial$-bar boundary condition has
 $\psi'(k,x,y)$
 as a solution. We call them {\bf the Special Contours of the second kind}. They are
  leaves of
 the  foliation on  2-torus
  given by the following equation:
$$(cke^{k\bar{z}}/\psi)d\bar{z}+(c\bar{k}e^{\bar{k}z}/\bar{\psi})dz=0$$ on the contour $\gamma$.
Here $c$ is real and nonzero.

The zeroes of $\psi'$ are singular points of this foliation. The manifold of
zeroes  $N_0=\{\psi'(k,x,y)=0\}$ is compact and two-dimensional
$N_0\subset {\mathbb C}\times {\mathbb R}^2=\{k,x,y\}$. The intersection of $N_0$ with every plane $k=const$
consists of finite number of isolated points. The surface $N_0$ projects into the $k$-plane.
 The image is a compact
set $N_0\rightarrow N^*\subset {\mathbb C}$. We know that $N^*$ is compact and does not cover $0$.
For all $k\in {\mathbb C}\setminus N^*$ our foliation on the
2-torus has no singular points. Our flow is analytic. Such flow is always homeomorphic to the straight line flow
with some
 ''rotation number'' $\rho(k)$ according to the famous classical results.
 So we have
$$
 \bar{\psi'}/\bar{k} e^{k\bar{z}}d\bar{z}+
 \psi'/ke^{k\bar{z}}dz=0.
$$
Let us introduce a real function
$$
 F=\left(\frac{1}{|k|^2}+\sum_j^n\left(a_j\frac{e^{l_j\bar{z}-\bar{l}_jz}}
{(k+l_j)(\bar{k}-\bar{l}_j)} +\bar{a}_j\frac{e^{-l_j\bar{z}+\bar{l}_jz}}{(k-l_j)(\bar{k}+\bar{l}_jz)}
\right)\right)e^{k\bar{z}+\bar{k}z}.
$$

Easy to check that $\partial F=2\psi'/ke^{\bar{k}z}$ and $\bar\partial
F=2\bar {\psi'}/\bar{k} e^{k\bar{z}}.$
 So for all $k\in {\mathbb C}\setminus N^*$ the contours are globally given by the equation
 $dF=0$ or $F=const$ in ${\mathbb R}^2$. For all values of $k\in {\mathbb C}$ the nonsingular levels $F=a$
 define trajectories which do not approach singularity. All these trajectories
 are either compact and homotopic to zero or topologically equivalent  in ${\mathbb R}^2$ to the straight
  line trajectory around the torus
 with some rotation number.  All components of the level $F=const$ approaching
   singular points  are either isolated singular points (the centers of foliation)
    or separatrices of the saddles (including
    degenerate saddles). So we proved the next statement:

    \begin{lem} Our foliation is  equivalent to Hamiltonian system (foliation)
    with
     Hamiltonian $F$ in every compact domain $E$ in the universal covering $E\subset {\mathbb R}^2$.
     \end{lem}
      As a corollary we conclude that\\
     1. All singular points are either centers or saddles
    (maybe degenerate).\\
    2. There are no limit cycles homotopic to zero in the torus, including
    limit cycles constructed from the pieces of separatrices from saddles to saddles.\\

    {\bf Is our foliation  globally topologically equivalent to the Hamiltonian foliation in the
    torus given by some multi-valued Hamiltonian (or closed one-form)?

     Can  limit cycles non-homotopic to zero
      exist for our system?} Let  such cycle $\gamma$ be given.
     It may be nonsingular
     or consists of several pieces of separatrices. Anyway, the value of function $F$ along
     $\gamma$ is constant, and
    the rotation number along $\gamma$ is rational.
According to our assumption, a family of non-closed open nonsingular
     trajectories $\gamma_1$ approaches $\gamma$ asymptotically for $t\rightarrow\infty$.
 We can always choose $\gamma_1$ such that
      $F(\gamma)\neq F(\gamma_1)$.

      There are two cases:

     {\bf Case 1:} The function $\psi'$ is unbounded along the closed contour $\gamma$. It is
     possible only for such contours that $F(\gamma)=0$ (see proof below).
      Any closed contour non-homotopic to zero
     in the torus with unbounded $\psi'$
     presents a limit cycle for our foliation --- see Example to the Case 1 and Fig 6 below:

     \vspace{1cm}

     {\bf Example to the case 1:} Let the functions $c,F$ are the same as above.
     We take following parameters:
     $$n=2,\quad l_1=1/2,\quad l_2=i/2,\quad a_1=a_2=0.2,\quad k=0.55i,$$
     $$c=1+0.4\cos(x)+0.4\cos(y).$$
     Making numerical calculation, we are coming to the limit cycles $\gamma$ where
      $b= [\gamma]\in H_1(T^2,{\mathbb Z})$ and
      $b$ is a basic cycle along the $y$-axis (imaginary). We have here
      $F(\gamma)= 0$:

\vspace{0.3cm}

    \begin{center}
{\color{Black}Fig 6}\\
 \vspace{2mm}
  \mbox{\epsfxsize=8cm\epsffile{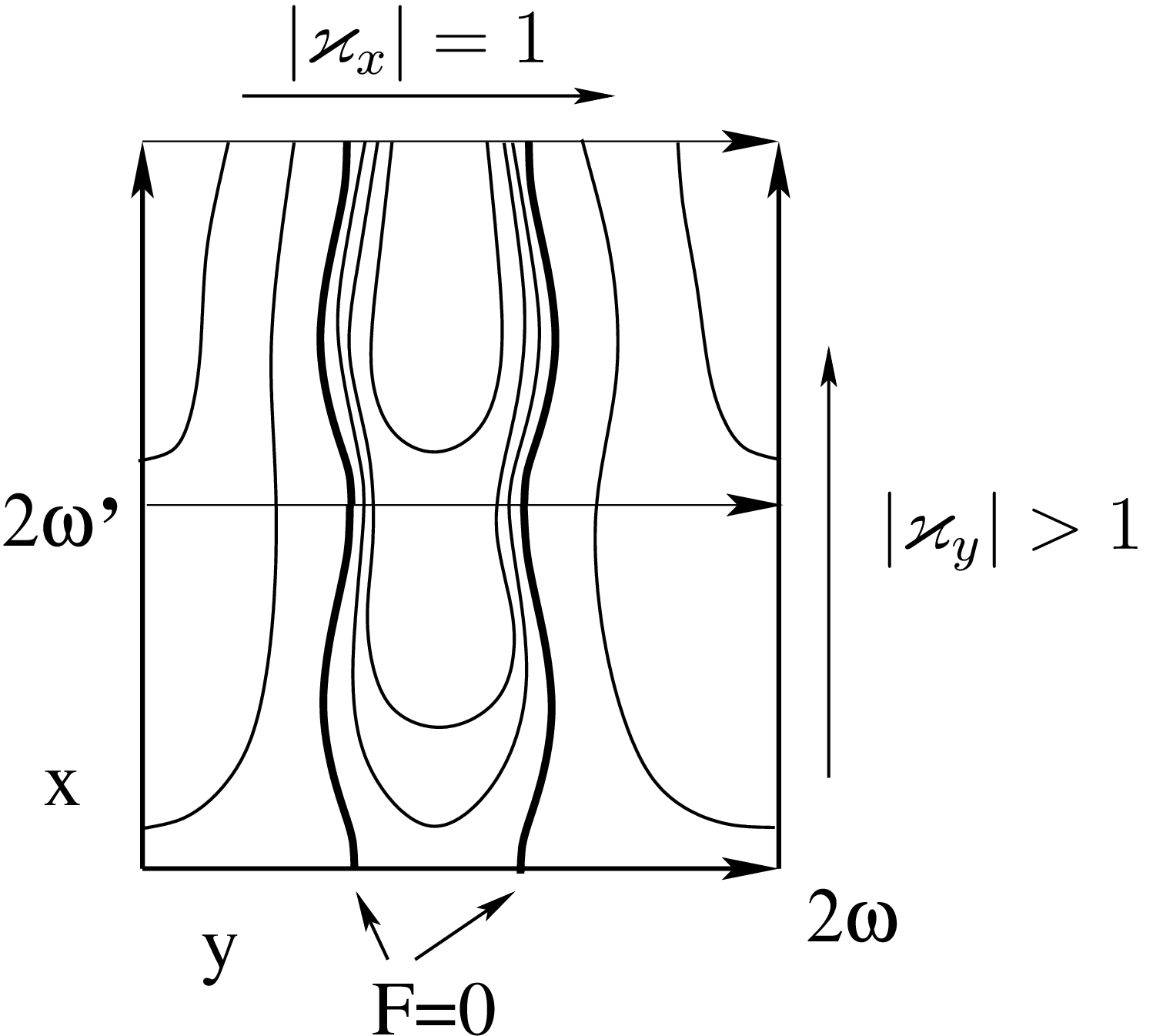}}\\
\end{center}

     \vspace{1cm}

Here the contours $F=0$ are closed and non-homotopic to 0.  Bloch multplier in the $y$-direction
 is different from
1. Therefore  differential $dF$ exponentially increases in the $y$-direction, and all contours
 $F=c$ tend to
the contour  $F=0$ as $y\rightarrow+\infty$

     {\bf  Case 2:} The function  $\psi'$ is bounded along the open  contours
     $\gamma_1$.\\
     {\bf According to the Lemma below it is always true if all connectivity components of the
       level $F= 0$ are compact in ${\mathbb R}^2$.} The
      rotation number is rational for the closed contour $\gamma$ non-homotopic to zero.
       So the trajectory $\gamma$
      corresponds to the shift $g: {\mathbb R}^2\rightarrow {\mathbb R}^2$ by the lattice
       vector with unitary
       multiplier. After the number of iterations $M$ we get rotation number for $g^M$ equal to 1.
      But the form $dF$ transforms
     as $g^*(dF)=\varkappa dF$ for all lattice shifts $g$ with real Bloch multipliers
      $g^*\psi'=\varkappa\psi',
     \varkappa\in {\mathbb R}$. We know that $F=const$
     along $\gamma_1$ and that  the form $dF$ is invariant under  the shift
     $g^M$ with unit multiplier $\varkappa=1$. So we see finally that $F(\gamma)=F(\gamma_1)$
      for all
     $\gamma_1$ approaching
     $\gamma$ if $\gamma$ is nonsingular. This is the contradiction for the nonsingular $\gamma$
     because $F$ is a smooth (even analytic) function. It is nonconstant
  along the transversal section to $\gamma$. It is true also for
      the singular
     $\gamma$ constructed from separatrices. To prove this statement we should use specific
     Bloch  properties
      of function $F$ written by the explicit formula, instead of $dF$ as above.
       In particular, we have to use its
      behavior under the lattice translations instead of the form $dF$.

      So {\bf we conclude that our foliation cannot have nonsingular (and singular)
     limit cycles nonhomotopic to zero and such that $F(\gamma)\neq 0$.}

 \begin{lem}Let
  $\hat{\gamma}$ be any open
 nonsingular trajectory such that topological closure of its projection $\gamma$ in the torus
  $T^2$ does not contain the levels $F=0$.
   Then $\psi'$
 is bounded along the
 contours $\gamma$. For every nonsingular periodic field $c\neq 0$ and $|k|$ big enough
 this condition is satisfied.
 \end{lem}

 Proof follows immediately from the formulas for the functions $F$ and $\psi'$ for  $|k|$
 big enough. The function
 $\psi'$ looks like $e^{k\bar{z}}$ multiplied by the bounded periodic function.
  So $|\psi'|\sim e^{{\rm Re}(k\bar{z})}$. The function $F$ looks like a
 bounded periodic function multiplied by the function $e^{2{\rm Re}(k\bar{z})}$. So the function
 $e^{{\rm Re}(k\bar{z})}$ is bounded along the levels $F=const\neq 0$.
  Therefore $\psi'$ is also bounded
 along this contour. For $|k|\rightarrow\infty$ big enough the asymptotic of function $F$
  looks like ($c/|k|^2)e^{k\bar{z}+\bar{k}z}$, so this function is separated from zero.
   Lemma is proved.

  We are coming to the following \\
  {\bf Conclusion}. There are only three possibilities for the leaves of this foliation
   for any $k\in {\mathbb C}$:\\
  1. There exists an open nonsingular
  trajectory $\gamma$ for which $\psi'(k,x,y)$ is a solution to the self-adjoint
  $\partial$-bar boundary problem in
 the domain $D\subset T^2$ like cylinder, $\hat{D}$ like strip  and $D_+$ or
 $D_-$
 like warped half-planes. One should choose only one  domain $D_+$ or $D_-$
  where $\psi'(k,x,y)/\sqrt{c}$
 has decay at infinity transversal to the boundary (see Fig 4)\\
 2. There exists a ''maximal'' cycle $\gamma$ homotopic to zero in the torus $T^2$ consisting of
 separatrices of the saddles such that $\gamma$ bounds the whole area of the torus  (see Fig 5).
 $\psi'$ gives solution to the self-adjoint boundary problem\\
 3. There exists a limit cycle $\gamma$ nonhomotopic to zero such that
  $F(\gamma)=0$ (see Fig 6). For such boundary contour $\psi'$ does not generate solution to
  the self-adjoint problem because it is has exponential growth along the boundary.\\

   This  picture is a natural 2D generalization of
 one-dimensional case
 where  functions  $f(x^*)=\psi(\gamma_j(x),x^*)$ give new levels in forbidden gaps of the periodic problem and describe
 an additional spectrum ---  Dirichlet spectrum at the interval $x^*\in [x,x+T]$ or ''border'' states in the half-line $R^+=[-\infty,x]$
  or $R^-=[x,+\infty]$ for the operator  $L=-\partial_{x^*}^2+u(x^*)$.

\end{document}